\documentclass[12pt]{article} 
\usepackage{amssymb,amsmath,color}

\numberwithin{equation}{section}
\usepackage[toc,page]{appendix}

\newcommand{\nc}{\newcommand}

\nc{\la}{\lambda} \nc{\alf}{\alpha} \nc{\La}{\Lambda} \nc{\ze}{\zeta}
\nc{\tht}{\theta} \nc{\T}{\Theta} \nc{\be}{\beta}  \nc{\eps}{\epsilon} 
\nc{\ga}{\gamma}  \nc{\De}{\Delta}  \nc{\G}{\Gamma}  \nc{\vphi}{\varphi}
\nc{\de}{\delta} \nc{\si}{\sigma}  \nc{\ka}{\kappa}   \nc{\Si}{\Sigma} 
\nc{\om}{\omega}  \nc{\qq}{\quad\quad}                \nc{\Om}{\Omega}
\nc{\nf}{\infty}   \nc{\dl}{\mathop{\smash{\cal L}}}  \nc{\black}{\rule{3mm}{3mm}}
\nc{\ra}{\rightarrow}    \nc{\ol}{\overline}        \nc{\und}{\underline} 
\nc{\beq}{\begin{equation}}  \nc{\eeq}{\end{equation}}  \nc{\pt}{\partial}  
   \nc{\dst}{\displaystyle}  \nc{\na}{\nabla} 
\nc{\nnb}{\nonumber}    \nc{\bs}{\backslash}        \nc{\mb}{\mathbb}   
\nc{\sn}{{\rm sn}\,} \nc{\cn}{{\rm cn}\,}     \nc{\dn}{{\rm dn}\,} \nc{\nin}{\noindent}
\nc{\ti}{\tilde}   \nc{\wti}{\widetilde}   \nc{\h}{\hat}  \nc{\wh}{\widehat}
\nc{\tpsi}{\wti{\psi}}   \nc{\tphi}{\wti{\phi}}  \nc{\tH}{\wti{H}} \nc{\Ai}{{\rm Ai}}

\newcounter{muni}
\newenvironment{remunerate}{\begin{list}{{\rm \arabic{muni}.}}
{\usecounter{muni}
\setlength{\leftmargin}{0pt}\setlength{\itemindent}{38pt}}}{\end{list}}

\nc{\brm}{\begin{remunerate}}   \nc{\erm}{\end{remunerate}}

\newtheorem{dfi}{Definition} \newtheorem{lem}{Lemma} 
\newtheorem{nth}{Proposition}  \newtheorem{nTh}{Theorem}
\newtheorem{conj}{Conjecture}
\nc{\stg}{\mathop{\smash{*}}}
\nc{\st}{\mathop{\smash{\delta}}}
\nc{\barr}{\begin{array}}   \nc{\earr}{\end{array}}   \nc{\dg}{\dagger}
\nc{\mtvb}{\mathversion{bold}}   \nc{\mtvn}{\mathversion{normal}}  \nc{\F}{f_{\eps}}

\begin{document} 

\begin{titlepage}

\date{\today}

\vspace{1cm}
\centerline{\huge\bf  Superintegrable models}

\vspace{5mm}
\centerline{\huge\bf on riemannian surfaces of revolution}

\vspace{7mm} 
\centerline{\huge\bf with integrals of any integer degree (I)}

\vskip 2.0truecm
\centerline{\large\bf  Galliano VALENT }

\vskip 2.0truecm
\centerline{ \it LPMP: Laboratoire de Physique Math\'ematique de Provence}
\centerline{\it 1 Avenue Marius Jouveau, 13090 Aix en Provence, France}
\nopagebreak

\vskip 2.5truecm

\begin{abstract} We present a family of superintegrable (SI) sytems living on a riemannian surface of revolution and which exhibits one linear integral and two integrals of any integer degree larger or equal to 2 in the momenta. When this degree is 2 one recovers a metric due to Koenigs.
 
The local structure of these systems is under control of  
a {\em linear} ordinary differential equation of order $n$ which is  homogeneous for even integrals and weakly inhomogeneous for odd integrals. The form of the integrals is explicitly given in the so-called ``simple" case (see definition \ref{simple}).
 
Some globally defined examples are worked out which live either in ${\mb H}^2$ or in ${\mb R}^2$.
\end{abstract}

\vskip 2truecm
\nin MSC 2010 numbers: 32C05, 81V99, 37E99, 37K25.

\end{titlepage}

\tableofcontents

\newpage\setcounter{section}{0}
\section{Introduction}
The possibility of integrable or SI  dynamical systems with integrals of degree larger than 4 in the momenta is such a difficult problem that the conjecture that they could not exist was put forward, see for instance \cite{bf}[p. 663]:

\begin{conj}[Kozlov and Fomenko] On the two dimensional sphere, there are no riemannian metrics whose geodesic flows are integrable by means of an integral of degree $n>4$ and do not admit integrals of degree $\leq 4$.
\end{conj} 

Up to now most of the integrable systems on riemannian surfaces, be explicit or not, do exhibit integrals of degree $\leq 4$ in the momenta. This apparent barrier opens the challenging question of the existence of integrable systems with integrals of strictly higher degree than $4$. 

The first example of an integrable system, on $S^2$, with integrals of sixth degree, emerging from astrophysics, was found by Gaffet \cite{Ga}. His results are described in a more attractive form in \cite{Ts1} and in \cite{bkm}. More recently Tsiganov gave a new example of this kind \cite{Ts2} on the two-sphere. In fact Kiyohara \cite{Ki} was the first to show the existence \footnote{However the explicit form of these metrics is not known.} of integrable systems with {\em globally defined} riemannian metrics having integrals of arbitrary integer degree which, furthermore, are Zoll metrics.

The study of SI models, generalizing Kepler and Hooke problems \cite{bbm}, has led also to interesting examples exhibiting integrals of any integer degree. For instance, considering the following generalization of the two dimensional Kepler problem
\beq
H=\frac 12\left(p_r^2+\frac{p_{\phi}^2}{r^2}\right)+\frac{k_1}{r}+\frac{k_2+k_3\,\cos(\la\phi)}{r^2\sin^2(\la\phi)}\qq\qq\la=\frac mn\eeq
the SI follows from Bertrand integral $f$ and the complex integrals of degree $m+n$:
\beq
\Phi^{(m,n)}=\left(\sqrt{2f}\,p_r-i(2f/r+k_1)\right)^m\left(\sqrt{2f}\,p_{\phi}\,\sin(\la\phi)+i(2f\,\cos(\la\phi)+k_3)\right)^n.\eeq
The linear (real) span of these integrals is four dimensional, but of course they must be functionally related.

The aim of this article is to construct SI systems defined on riemannian surfaces of revolution, starting from the framework laid down by Matveev and Shevchishin \cite{ms} for cubic integrals. Their analysis, as we will show later on, can be  generalized to integrals of any degree, starting from degree 2. In this last simple case it was proved in \cite{VaK} that one recovers Koenigs metrics \cite{Ko} studied and generalized in \cite{kkmw}. Koenigs systems exhibit the following integrals
\beq
H \qq P_y \qq S_1 \qq S_2\eeq
where $\ H,\,S_1,\,S_2$ are of second degree in the momenta. In our generalization the integrals $S_1$ and $S_2$ will be of any integer degree $\geq 3$. Of course these four quantities, as we will show explicitly, are algebraically related.

Since the analysis required for even and odd degrees integrals display some differences we have divided the article in two Parts. 

In Part I we consider the case of integrals $S_1$ and $S_2$ of degree $2n\geq 2$. Their local structure is constructed and shown to be determined by a linear and homogeneous ODE of order $n$. Globally defined examples are given either on $M={\mb H}^2$ or on $M={\mb R}^2$.

In Part II the case of integrals $S_1$ and $S_2$ of degree $2n+1\geq 3$ is considered. Their local structure is constructed and shown to be determined by a linear and weakly inhomogeneous ODE of order $n$. Globally defined examples give rise to the same manifolds as in Part I.
The last Section is devoted to some concluding remarks.

\part{Integrals of even degree in the momenta}
\setcounter{section}{1}

\vspace{4mm} 
We will consider the cases for which the observables have for degree in the momenta $\sharp(Q)=2n$ 
where $n\geq 2$. The case $n=1$ is marginal since the  corresponding SI models were discovered by Koenigs in \cite{Ko} and generalized in \cite{kkmw}.

Taking for hamiltonian
\beq
H=\Pi^2+a(x)\,P_y^2\qq\qq \Pi=a(x)\,P_x\eeq
we have the obvious result:
\begin{nth}
The system $(H,P_y)$ is integrable in Liouville sense.
\end{nth}

\nin{\bf Proof:} We have $\{H,P_y\}=0$ and since $H$ and $P_y$ are generically independent the proposition follows. $\quad\Box$

This dynamical system will become SI  if we can construct at least one more, generically independent integral, of degree $2n$ in the momenta. To this aim let us first define
\beq
G=\sum_{k=0}^n\,A_k\,H^k\,P_y^{2(n-k)}\qq\qq A_n\neq 0\qq\qq\qq \sharp(G)=2n \qq n\geq 1.
\eeq
Since $A_n$ cannot vanish, we will set $A_n=1$. Therefore the function $G$ defines a string of $n$ real constants: $\ (A_0,\,A_1,\cdots,A_{n-1})$.
 
The definition of the two observables $(S_1,\,S_2)$ by  
\beq\label{S1S2}
S_1=Q_1+y\,G \qq\qq S_2=Q_2+y\,Q_1+\frac{y^2}{2}\,G\qq\qq \sharp(Q_1)=\sharp(Q_2)=2n,
\eeq
entail the relations
\beq
\{P_y,S_1\}=G\qq\qq\{P_y,S_2\}=S_1.
\eeq
These observables will become integrals if we impose:

\begin{nth}\label{Sint1} The observables $S_1$ and $S_2$ are integrals iff
\beq\label{condSint}
\{H,Q_1\}+2a\,P_y\,G=0\qq\qq \{H,Q_2\}+2a\,P_y\,Q_1=0.
\eeq
\end{nth}

\nin{\bf Proof:} The first relation is nothing but  
$\{H,S_1\}=0$ and the second one is $\{H,S_2\}=0$. 
$\quad\Box$

Let us observe that we get in fact two (maximally) SI systems which are
\beq
{\cal I}_1=\{H,\,P_y,\,S_1\} \qq  \&  \qq {\cal I}_2=\{H,\,P_y,\,S_2\}.
\eeq
For further use let us define the sets
\[{\cal S}^p_n=\{p,p+1,\ldots,n\}\qq\qq 0\leq p\leq n\]
as well as the Pochammer symbols
\[ (z)_0=1\qq \forall n\geq 1 \qq (z)_n=z(z+1)\cdots(z+n-1).\]

\section{The local structure of the first integral}
The first step will be

\begin{nth}\label{sys0} The  most general form of $\ Q_1$ being  
\beq
Q_1=\sum_{k=1}^n\,b_k(x)\,\Pi^{2k-1}\,P_y^{2(n-k)+1}\qq\qq \sharp(Q_1)=2n \qq n\geq 2,
\eeq
the constraint $\ \{H,S_1\}=0$ is equivalent to the differential system \footnote{A prime stands 
for a derivation with respect to the variable $x$.}:
\beq
\left\{\barr{l} 0=\frac 12\,a'\,b_1-F\\[4mm]\dst  
b'_k=(k+\frac 12)\,a'\,b_{\,k+1}-\frac{D_a^k\,F}{k!}
\qq k \in \{1,\,2,\ldots,\,n-1\}\\[4mm]\dst 
b'_n=-\frac{D_a^n\,F}{n!},\earr\right.\eeq
where $\ F(a)=\dst\sum_{k=0}^n\,A_k\,a^k$.
\end{nth}

\nin{\bf Proof:} The first relation in (\ref{condSint}) is
\beq\label{eqP1}
\{H,Q_1\}+2a\,P_y\,G=0.\eeq
Expanding the left hand side leads to
\[\barr{l}\dst \sum_{k=1}^{n-1}\,\Big(b'_k-(k+1/2)\,a'\,b_{k+1}+\frac{D_a^k\,F}{k!}\Big)\,a\,\Pi^{2k}\,P_y^{2(n-k)+1}\\[4mm]\dst \hspace{3cm}
+\Big(F-\frac 12\,a'\,b_1\Big)a\,P_y^{2n+1}+\Big(b'_n+\frac{D_x^n\,F}{n!}\Big)a\,\Pi^{2n}\,P_y=0\earr\]
which proves the Proposition. $\quad\Box$

To reduce this system to a tractable form we will use now, instead of the coordinate $x$, the coordinate $a$. 
This is legitimate since our considerations are purely local. The hamiltonian becomes
\beq\label{H2}
H=\Pi^2+a\,P_y^2\qq\qq \Pi=\frac{a}{\dot{x}}\,P_a.\qq\qq \eeq
Transforming the equations in Proposition \ref{sys0} gives

\begin{nth} The constraint $\{H,S_1\}=0$ is equivalent to the differential system \footnote{A dot stands for  a derivation with respect to the variable $a$.}:
\beq\label{eqP3}\left\{\barr{ll} (a) & \dst\qq b_1=2F\ \dot{x}\\[2mm] 
(b) & \qq\dst \dot{b}_k=(k+\frac 12)\,b_{k+1}-\frac{D_a^k\,F}{k!}\ \dot{x} \qq k \in \{1,\,2,\ldots,\,n-1\}\\[4mm]
(c) & \qq\dst \dot{b}_n=-\frac{D_a^n\,F}{n!}\ \dot{x}=-\dot{x}\earr\right.\eeq
\end{nth}
Let us notice that we have $n+1$ equations for $n$ unknown functions $\ \Big(b_k(a),\ k\in{\cal S}^1_n\Big)$.

The definition will be useful:

\begin{dfi} The linear differential operator ${\rm Op}_n[F]$ is defined as
\beq
{\rm Op}_n[F]=\sum_{s=0}^n\,\frac{F^{(n-s)}}{(n-s)!}\frac 1{(1/2)_s}\,D_a^s.\eeq
\end{dfi}
The Leibnitz formula gives for it the following property:
\beq\label{idop}
{\rm Op}_n[FG]=\sum_{s=0}^n\frac{F^{(n-s)}}{(n-s)!}\ {\rm Op}_s[G].\eeq

We are now in position to solve the system for the $b_k$:

\begin{nth}[Linearizing ODE]\label{Q1bk} The local structure of $Q_1$ 
is given by
\beq
Q_1=\sum_{k=1}^n\,b_k\,\Pi^{2k-1}\,P_y^{2(n-k)+1},
\eeq
with 
\beq\label{bk}
\forall k\in{\cal S}^1_n \qq\qq 
b_k[F]=\sum_{s=1}^k\,\frac{F^{(k-s)}}{(k-s)!}\,\frac{D_a^s\,x}{(1/2)_s}={\rm Op}_k[F]x-\frac{F^{(k)}}{k!}\,x\eeq
where $x$ is a solution of the {\em linear} and {\em homogeneous} ODE of order $n\geq 2$:
\beq\label{odeP3}
{\rm Op}_n[F]x=0.\eeq
\end{nth}

\nin{\bf Proof:} To determine the functions $b_k$ we need to solve the differential system (\ref{eqP3}): we will proceed recursively. The relation (\ref{eqP3})(a) gives
\[b_1=F\ \frac{D_a\,x}{1/2}\]
in agreement with (\ref{bk}) for $k=1$. Let it be supposed that the relation for $b_k$ in (\ref{bk}) is true  for any $k<n$ and let us prove that it is also 
true for $k+1\leq n$. To this aim we will start 
from relation (\ref{eqP3})(b) written
\[(k+1/2)b_{k+1}=\dot{b}_k+\frac{D_a^k\,F}{k!}\,\dot{x}=
\sum_{s=1}^k\,\frac{D_a^{k+1-s}\,F}{(k-s)!}\,
\frac{D_a^s\,x}{(1/2)_s}+
\sum_{s=0}^k\,\frac{D_a^{k-s}\,F}{(k-s)!}\,\frac{D_a^{s+1}\,x}{(1/2)_s}.\]
In the second sum let us shift $s \to s'=s+1$ so that
\[(k+1/2)b_{k+1}=\sum_{s=1}^{k}\,\frac{D_a^{k+1-s}\,F}{(k-s)!}\,\frac{D_a^s\,x}{(1/2)_s}+
\sum_{s'=1}^{k+1}\,\frac{D_a^{k+1-s'}\,F}{(k+1-s')!}\,\frac{D_a^{s'}\,x}{(1/2)_{s'-1}}\]
\[=F\,\frac{D_a^{k+1}\,x}{(1/2)_k}
+\sum_{s=1}^k\left[\frac 1{(1/2)_{s-1}}+\frac{k+1-s}{(1/2)_s}\right]\frac{D_a^{k+1-s}\,F}{(k+1-s)!}\,D_a^s\,x.\]
The relations
\[\frac 1{(1/2)_{s-1}}+\frac{k+1-s}{(1/2)_s}=\frac{k+1/2}{(1/2)_s}\qq\qq (1/2)_{k+1}=(k+1/2)\,(1/2)_k\]
give for final result
\[b_{k+1}=\sum_{s=1}^{k+1}\,\frac{D_a^{k+1-s}F}{(k+1-s)!}\,\frac{D_a^s\,x}{(1/2)_s}\]
which concludes the recurrence proof of the relation (\ref{bk}) for $b_k$. 

We are left with relation (\ref{eqP3})(c) which integrates up to $\ b_n=-A_n\,(x-x_0)$ and this last relation {\em must} agree with the 
$b_n$ obtained from (\ref{eqP3})(b) for $k=n-1$. This produces the linear ODE
\[\sum_{s=1}^n\,\frac{D_a^{n-s}\,F}{(n-s)!}\,\frac{D_a^s\,x}{(1/2)_s}+\frac{D_a^{n}\,F}{n!}\,(x-x_0)=0.\]
We can set $\,x_0=0$ because the metric does depend solely on $\dot{x}$ and this proves (\ref{odeP3}).
$\quad\Box$

\vspace{3mm}
\nin{\bf Remark:} Let us observe that the relation (\ref{bk}) is valid for {\em any} choice of $F$.

Let us define:
\begin{dfi}\label{simple} We will say that $F$ is {\em simple} if all of its zeroes are simple, with symbol $\wh{F}$:
\[\wh{F}(a)\equiv \sum_{k=0}^n\,A_k\,a^k=\prod_{i=1}^n(a-a_i)\qq\qq A_n\in{\mb R}\bs\{0\}.\]
\end{dfi} 
Then we have 

\begin{nth}\label{propode} For a simple $F$, if one takes
\beq\label{xgene}
x=\sum_{i=1}^n\,\frac{\xi_i}{\sqrt{\De_i}}\qq\qq \De_i=\eps_i(a-a_i)\qq\qq \eps_i^2=1 \qq a_i\in{\mb R} \eeq
where the $\xi_i$ are $n$ arbitrary real parameters, the relation
\beq\label{solode}
{\rm Op}_n[\wh{F}]\,x=\sum_{i=1}^n\,\xi_i\,\frac{\De_i^{1/2}}{n!}\,D_a^n\left(\frac {\wh{F}}{\De_i}\right)\eeq
implies that $x$ is the general solution of the ODE (\ref{odeP3}).
\end{nth}

\nin{\bf Proof:} The linearity of the ODE allows to check  term by term 
\[\forall i\in{\cal S}^1_n \qq\qq x_i=\De_i^{-1/2}\qq\qq \De_i=\eps_i(a-a_i)\qq\qq \qq \eps_i^2=1,\]
for which 
\beq\label{dersx}
\forall\,s\in{\mb N} \qq\qq \frac{D_a^s\, x_i}{(1/2)_s}=(-\eps_i)^s\,\De_i^{-s-1/2}.\eeq
Hence 
\[{\rm Op}_n[\wh{F}]\,x_i=\De_i^{1/2}
\sum_{s=0}^n\,\frac{D_a^{n-s}\,\wh{F}}{(n-s)!}\,(-\eps_i)^s\,\De_i^{-s-1}\]
so that using
\[\forall\,s\in{\mb N} \qq\qq (-\eps_i)^s\,\De_i^{-s-1}=\frac 1{s!}\,D_a^s\,(\De_i^{-1})\]
and Leibnitz formula we end up with
\[
{\rm Op}_n[\wh{F}]\,x_i=
\De_i^{1/2}\ \frac{D_a^n}{n!}\left(\frac {\wh{F}}{\De_i}\right).\]
Adding all the terms gives (\ref{solode}). The proposition follows since each term in this sum does vanish.
$\quad\Box$

\vspace{3mm}
\nin{\bf Remarks:} 

\brm
\item The previous formula is in fact valid for
\beq\label{allk}
\forall k\,\in\,{\cal S}^0_n: \qq  {\rm Op}_k[\wh{F}]\,x_i=
\De_i^{1/2}\,\frac{D_a^k}{k!}\left(\frac {\wh{F}}{\De_i}\right),\eeq
which does vanish only for $k=n$. 
\item The forms taken by $x(a)$ when $F$ is not simple are given in Appendix A.
\erm
As a bonus the metrics of constant scalar curvature are excluded:
 
\begin{nth}\label{ccurv} The metric
\beq
g=\frac{\dot{x}^2}{a^2}\,da^2+\frac{dy^2}{a}\eeq
is never of constant curvature if  $x(a)$ is a solution of the ODE (\ref{odeP3}). An embedding in ${\mb R}^{2,1}$ is given by
\beq
g=dX^2+dY^2-dZ^2\eeq
where
\beq
X=\frac y{\sqrt{a}}\qq Y-Z=-\frac 1{\sqrt{a}} \qq 
Y+Z=\frac{y^2}{\sqrt{a}}+2\int\,\frac{\dot{x}^2}{\sqrt{a}}\,da.\eeq
\end{nth}

\nin{\bf Proof:}
We have seen that the scalar curvature is
\beq\label{scalR}
2\,R=-\frac{2a\,\ddot{x}+\dot{x}}{(\dot{x})^3}
\eeq
so if we take $R$ to be a constant, defining $\dst u=\frac 1{\dot{x}^2}$ we have to solve
\[a\,\dot{u}-u=2R\qq\Longrightarrow\qq u=K\,a-2R.\]
According to the value of the integration constant $K$, and omitting an additive constant, we have
\[K=0\quad\Longrightarrow\quad x=\pm\frac{a}{\sqrt{-2R}},\qq\qq K\neq 0\quad\Longrightarrow\quad  
x=\pm\frac 2K\,\sqrt{K\,a-2R}\]
and both functions are never solutions of (\ref{odeP3}).
 The embedding formulas are easily checked. $\quad\Box$

For the next step we need 
\begin{lem} One has the following identity
\beq\label{idkl}
k\leq l: \qq \sum_{s=k}^l\frac{(-1)^s}{(l-s)!\,(s-k)!}=(-1)^k\,\de_{kl}.\eeq
\end{lem}

\nin{\bf Proof:} For $k=l$ this sum is just $\,(-1)^k$. For $\,k<l$ defining $\,N=l-k$ and $t=s-k$ 
we have
\[\sum_{s=k}^l\frac{(-1)^s}{(l-s)!\,(s-k)!}= \sum_{t=0}^N\frac{(-1)^{k+t}}{t!\,(N-t)!}=
\frac{(-1)^k}{N!}\sum_{t=0}^N (-1)^t\,\binom{N}{t}=0\]
by the binomial theorem.$\quad\Box$

Let us first compute the coefficients $b_k$:
\begin{nth} For a simple $F$ we have
\beq
Q_1=\sum_{k=1}^n\,b_k[\wh{F}]\,\Pi^{2k-1}\,
P_y^{2(n-k)+1},
\eeq
with
\beq\label{formbk1}
\forall k \in {\cal S}^1_n \qq\qq b_k[\wh{F}]=-\sum_{i=1}^n\,\frac{\eps_i\,\xi_i}{\sqrt{\De_i}}\,\frac{D_a^{k-1}}{(k-1)!}\left(\frac{\wh{F}}{\De_i}\right).
\eeq
\end{nth}

\nin{\bf Proof:} Relation (\ref{bk}) 
\[
b_k[\wh{F}]=\sum_{s=1}^k\,\frac{\wh{F}^{(k-s)}}{(k-s)!}\,\frac{D_a^s x}{(1/2)_s}\qq\qq x=\sum_{i=1}^n\,\frac{\xi_i}{\sqrt{\De_i}}\]
and the identity (\ref{dersx}) give first
\[
b_k[\wh{F}]=\sum_{i=1}^n \,\frac{\xi_i}{\sqrt{\De_i}}\sum_{s=1}^k\frac{\wh{F}^{(k-s)}}{(k-s)!}
(-\eps_i)^s\,\De_i^{-s}.\]
The change of index $s=t+1$ gives
\[
b_k[\wh{F}]=-\sum_{i=1}^n \,\frac{\eps_i\,\xi_i}{\sqrt{\De_i}}\sum_{t=0}^{k-1}\frac{\wh{F}^{(k-1-t)}}{(k-1-t)!}
(-\eps_i)^{t}\,\De_i^{-t-1}.\]
Using the identity
\[
(-\eps_i)^t\,\De_i^{-t-1}=\frac{D_a^t\De_i^{-1}}{t!}\]
and Leibnitz formula, we are led to (\ref{formbk1}).
$\quad\Box$

Using this form of $Q_1$ it is rather difficult to obtain  $Q_2$. To solve this problem we need to transform $Q_1$ according to
\beq
Q_1=\sum_{k=1}^n\,b_k[F]\,\Pi^{2k-1}\,P_y^{2(n-k)+1}=\sum_{k=1}^n\,\wti{b}_k[F]\,H^{n-k}\,\Pi\,P_y^{2k-1}.
\eeq
Let us determine the new coefficients $\,\wti{b}_k$:

\begin{nth}\label{bt9} In general we have
\beq\label{bktilde}
\forall k\in {\cal S}^1_n: \qq\qq \wti{b}_k[F]=\sum_{s=1}^k{n-s \choose k-s}\,(-a)^{k-s}\,b_{n-s+1}[F]\eeq
and in the simple case \footnote{The symbol $\si^i_{k-1}$ is defined in Appendix B.}
\beq\label{bktsimple}
\forall k\in {\cal S}^1_n: \qq\qq \wti{b}_k[\wh{F}]=(-1)^k\,\sum_{i=1}^n\frac{\xi_i}{\sqrt{\De_i}}\,\si^i_{k-1}.
\eeq
\end{nth}

\nin{\bf Proof:} If, in the first form of $Q_1$, one uses $\Pi^2=H-a\,P_y^2$ and interchanges the summations order, one gets the relation (\ref{bktilde}). Using formula (\ref{formbk1}) we have
\[ b_{n+1-s}[\wh{F}]=-\frac 1{(n-s)!}\sum_{i=1}^n\,\frac{\eps_i\xi_i}{\sqrt{\De_i}}\,D_a^{n-s}\left(\frac{\wh{F}}{a-a_i}\right)
\qq\qq s\in {\cal S}^1_n.\]
Expanding the term inside the bracket using relation (\ref{sfr1}) one has
\[b_{n+1-s}[\wh{F}]=-\frac{1}{(n-s)!}\sum_{i=1}^n\,\frac{\xi_i}{\sqrt{\De_i}}\,
\sum_{l=1}^s\,(-1)^{l-1}\,\si^i_{l-1}\,\frac{(n-l)!}{(s-l)!}\,a^{s-l},\]
and inserting this formula in the definition, given above, of $\wti{b}_k$ we get
\[
\wti{b}_k[\wh{F}]=(-1)^{k+1}\,\sum_{i=1}^n\,\frac{\xi_i}{\sqrt{\De_i}}\,
\sum_{s=1}^k\sum_{l=1}^s\,\frac{(-1)^{l-1+s}}{(k-s)!}\,
\frac{(n-l)!}{(n-k)!}\,\frac{a^{k-l}}{(s-l)!}\,\si^i_{l-1}.\]
Reversing the first and the second summations we end up with
\[
\wti{b}_k[\wh{F}]=(-1)^{k+1}\,\sum_{i=1}^n\,\frac{\xi_i}{\sqrt{\De_i}}\,
\sum_{l=1}^k(-1)^{l-1}\si^i_{l-1}\,a^{k-l}\frac{(n-l)!}{(n-k)!}\sum_{s=l}^k\frac{(-1)^s}{(k-s)!(s-l)!}
\]
and the identity (\ref{idkl}) concludes the proof. 
$\quad\Box$

It is interesting, in order to check the result obtained in Proposition \ref{bt9}, to write down  the differential system for the $\wti{b}_k$. We have:

\begin{nth}\label{eqbk0} Defining
\beq
\forall k \in\,{\cal S}^1_n: \qq 
\wti{b}_k[F]=(-1)^k\,\be_k[F],\eeq 
for any choice of $F$ the relations 
\beq\label{sysbe}\barr{ll}
  & \qq \dot{\be}_1=\dot{x}\\[4mm]\dst 
1\leq k\leq n-1: & \qq\dot{\be}_{k+1}=-a\,\dot{\be}_k-\frac 12\,\be_k+\si_k\,\dot{x}\\[4mm] 
k=n:  & \qq 0=-a\,\dot{\be}_n-\frac 12\,\be_n+\si_n\,\dot{x},\earr\eeq
ensure the conservation of $S_1$. For a simple $F$ the formula obtained for $\,\wti{b}_k[\wh{F}]\,$ in Proposition \ref{bt9} is indeed a solution of this system. 
  
\end{nth}

\nin{\bf Proof:} A routine computation leads to 
\[\barr{l}\dst 
0=\{H,Q_1\}+2a\,P_y\,G=-\frac{2a}{\dot{x}}
\left\{\sum_{k=0}^{n-1}\,(-1)^k\,\dot{\be}_{k+1}\,H^{n-k}\,P_y^{2k+1}\right.\\[4mm]\dst \left. \hspace{6cm}+\sum_{k=1}^n\,(-1)^k\left(a\dot{\be}_k+\frac 12\,\be_k-(-1)^kA_{n-k}\dot{x}\right)\,H^{n-k}\,P_y^{2k+1}\right\}\earr\]
which gives (\ref{sysbe}) using the relation (\ref{sik}).

Let us now check that the formula
\[\forall k\in\,{\cal S}^1_n \qq\quad \be_k[\wh{F}]=\sum_{i=1}^n\frac{\xi_i}{\sqrt{\De_i}}\,\si_{k-1}^i\]
proved in Proposition \ref{bt9} does solve this system.

For $k=1$ we have
\[\be_1=\sum_{i=1}^n\frac{\xi_i}{\De_i}\,\si_0^i=\sum_{i=1}^n\frac{\xi_i}{\sqrt{\De_i}}=x.\]
For $k\in{\cal S}^2_n$, using the relation (\ref{id1sfr}), we have
\[-a\dot{\be}_k-\frac 12\be_k+\si_k\dot{x}=\sum_{i=1}^n\frac{\eps_i\xi_i}{2\De_i^{3/2}}(a_i\si^i_{k-1}-\si_k)=-\sum_{i=1}^n\frac{\eps_i\xi_i}{2\De_i^{3/2}}\,\si^i_k.\]
If $k\in\{1,2,\ldots,n-1\}$ we do recover $\dot{\be}_{k+1}$, while for $k=n$ the result vanishes. 
$\quad\Box$

\vspace{3mm}
\nin{\bf Remarks:}
\brm
\item Using symbolic computation we checked the conservation of $S_1=Q_1+y\,G$ using for $Q_1$ its form (\ref{bktsimple}) for $n=2$ and $n=3$.
\item An interesting exercise, left for the reader, is to derive the differential system (\ref{sysbe}) from the relation (\ref{bktilde}) and the differential system for the coefficients $b_k$.
\erm

As we will see now this last form of $Q_1$ will allow for a simple construction of $Q_2$.

\section{The local structure of the second integral}
Let us proceed with $\ Q_2$. From Proposition \ref{Sint1} we need to solve 
\[ \{H,Q_2\}+2a\,P_y\,Q_1=0. \]

Let us prove
\begin{nth}\label{propQ2} For a simple $F$, the observable 
\beq\label{Q2final}
Q_2=\sum_{k=1}^n\,\wti{c}_k[\wh{F}]\,H^{n-k}\,P_y^{2k}\eeq
is given by 
\beq\label{resQ2}
\forall k\in {\cal S}^1_n \quad   
\wti{c}_k[\wh{F}]=\frac{(-1)^{k+1}}{2}\,\left(\sum_{i=1}^n\frac{\xi_i^2}{\De_i}\ \si^i_{k-1}+
\sum_{i\neq j=1}^n\,\frac{\xi_i\,\xi_j}{\sqrt{\De_i\,\De_j}}\Big(\si^{ij}_{k-1}+a\,\si^{ij}_{k-2}\Big)\right)
\eeq
where the $\si^{ij}_{k}$ are defined in Appendix B. 
\end{nth}

\nin{\bf Proof}: An elementary computation gives
\[\{H,Q_2\}+2a\,P_y\,Q_1=
2a\,\Pi\sum_{k=1}^n\Big(\frac{D_a \wti{c}_k}{\dot{x}}+\wti{b}_k\Big)H^{n-k}\,P_y^{2k}.\]
Since we are working locally, this is equivalent to
\[\forall k\in {\cal S}^1_n \qq\qq D_a\,\wti{c}_k=-\wti{b}_k\,\dot{x} \]
or explicitly
\[D_a\,\wti{c}_k=\frac{(-1)^k}{2}\,\sum_{i=1}^n\,\frac{\xi_i}{\sqrt{\De_i}}\,\si^i_{k-1}\sum_{j=1}^n\frac{\eps_j\,\xi_j}{(\De_j)^{3/2}}.\]
Expanding into
\[
D_a\,\wti{c}_k=\frac{(-1)^k}{2}\,\left(\sum_{i=1}^n\frac{\eps_i\,\xi_i^2}{(a-a_i)^2}\,\si^i_{k-1}+
\sum_{i\neq j=1}^n\,\frac{\xi_i\,\xi_j\,\eps_j}{\sqrt{\De_i}\,(\De_j)^{3/2}}\,\si^i_{k-1}\right)\]
and integrating up to \footnote{The integration constants  can be omitted since they would add trivially conserved terms.}
\[
\wti{c}_k=\frac{(-1)^k}{2}\,\left(-\sum_{i=1}^n\frac{\xi_i^2}{\De_i}\,\si^i_{k-1}+
2\sum_{i\neq j=1}^n\,\frac{\xi_i\,\xi_j(a-a_i)}{(a_i-a_j)\sqrt{\De_i\,\De_j}}\,\si^i_{k-1}\right).
\]
The second piece can be written
\[
2\sum_{i>j=1}^n\,\frac{\xi_i\,\xi_j}{\sqrt{\De_i\,\De_j}}
\frac{\Big(\si^i_{k-1}(a-a_i)-\si^j_{k-1}(a-a_j)\Big)}{a_i-a_j}.\]
Using the relation (\ref{id1sfr}) gives
\[ 
\si^i_{k-1}(a-a_i)-\si^j_{k-1}(a-a_j)=a(\si^i_{k-1}-\si^j_{k-1})+\si^i_k-\si^j_k\]
and thanks to (\ref{id3sfr}) we obtain (\ref{resQ2}).
$\quad\Box$

\vspace{2mm}
\nin{\bf Remark}: For $k=1$ we have $D_a\wti{c}_1=-\wti{b}_1\,\dot{x}=x\,\dot{x}$ which integrates up to $\dst\wti{c}_1=\frac{x^2}{2}$. It does agree with formula (\ref{resQ2}) since we have $\si^{ij}_0=1$ and $\si^{ij}_{-1}=0$.

Let us add an important algebraic relation:

\begin{nth} 
The integrals $S_1$ and $S_2$ are algebraically related by
\beq\label{idQ1}
S_1^2-2G\,S_2=A_n^2\,\sum_{k,l=1}^n\,{\cal Q}_{kl}\,H^{2n-k-l}\,P_y^{2(k+l)}\eeq
where
\beq\label{defqkl}
{\cal Q}_{kl}=(-1)^{k+l+1}\,\sum_{i=1}^n\eps_i\,\xi_i^2\,\si^i_{k-1}\,\si^i_{l-1}.
\eeq 

\end{nth}

\nin{\bf Proof:} We have first
\[X\equiv S_1^2-2GS_2=Q_1^2-2GQ_2.\]
Expanding this expression in powers of $H$ and of $P_y$ and upon use of the identities (\ref{id1sfr}) and (\ref{idQ}) leads, after some hairy computations to the given formula.$\quad\Box$

Summarizing the results obtained up to now we have:

\begin{nTh}\label{thun} The hamiltonian
\beq
H=\Pi^2+a\,P_y^2\qq\qq \Pi=\frac{a}{\dot{x}}\,P_a\qq\qq a>0\eeq
for a {\em simple} $F=\wh{F}$ and
\beq
x=\sum_{i=1}^n\,\frac{\xi_i}{\sqrt{\De_i}}\qq \De_i=\eps_i(a-a_i)\qq\qq \eps_i^2=1 \qq\qq \xi_i\,\in\,{\mb R},\eeq
exhibits two integrals
\beq
S_1=Q_1+y\,G \qq\qq S_2=Q_2+y\,Q_1+\frac{y^2}{2}\,G
\eeq
where
\beq
G=\sum_{k=0}^n\,A_{n-k}\,H^{n-k}\,P_y^{2k}\quad Q_1=\sum_{k=1}^n\,\wti{b}_k\,H^{n-k}\,\Pi\,P_y^{2k-1} 
\quad   Q_2=\sum_{k=1}^n\,\wti{c}_k\,H^{n-k}\,P_y^{2k}
\eeq
and
\beq\label{btkctk1}
\forall k \in {\cal S}^1_n:\quad \left\{\barr{l}\dst 
\wti{b}_k=(-1)^k\,\sum_{i=1}^n\,\frac{\xi_i}{\sqrt{\De_i}}\,\si^i_{k-1} \\[6mm]\dst 
\wti{c}_k=\frac{(-1)^{k+1}}{2}\left(\sum_{i=1}^n\frac{\xi_i^2}{\De_i}\,\si^i_{k-1}+
\sum_{i\neq j=1}^n\,\frac{\xi_i\,\xi_j}{\sqrt{\De_i\,\De_j}}
\Big(\si^{ij}_{k-1}+a\,\si^{ij}_{k-2}\Big)\right).\earr\right.
\eeq
These two integrals generate  two maximally SI systems:
\beq
{\cal I}_1=\{H,\,P_y,\,S_1\}\qq\qq \mbox{and}\qq\qq {\cal I}_2=\{H,\,P_y,\,S_2\}.
\eeq
\end{nTh}

\nin{\bf Proof:} We just need to check the functional independence of the integrals. Let us define
\[J_1=dH\wedge dP_y\wedge dS_1\qq\qq J_2=dH\wedge dP_y\wedge dS_2.\]
We have 
\[J_1=dH\wedge dP_y\wedge dS_1=y\,dH\wedge dP_y\wedge \,dQ_1+Q_1\frac{\pt H}{\pt P_x}\,dP_x\wedge dP_y\wedge dy\]
which cannot vanish everywhere due to the last term. Differentiating the identity (\ref{idQ1}) we have
\[d(S_1)^2-2G\,dS_2=2dG\,S_2+dX(H,P_y)\]
which implies that
\[2G\,J_2=dH\wedge dP_y\wedge (2G\,dS_2)=2S_1\,J_1\]
does not vanish everywhere. $\quad\Box$

\section{Cascading}
Let us first introduce some notations which make explicit the degree $2n$ of the integrals. Notice that we have to bring back the coefficient $A_n$ which was taken to be 1. 
We have first
\beq
G^{(n)}=\sum_{k=0}^n\,A_k\,H^k\,P_y^{2(n-k)}\qq 
F^{(n)}=\sum_{k=0}^n\,A_k\,a^k=A_n\prod_{l=1}^n(a-a_l)\eeq
and similarly for the symmetric functions of the roots: \beq
\si_k^{(n)}\qq\qq \si^{i\,(n)}_{k-1} \qq\qq 
\si^{ij\,(n)}_{k-2}.
\eeq
The hamiltonian is
\beq
H^{(n)}=(\Pi^{(n)})^2+a\,P_y^2\qq\qq\Pi^{(n)}=\frac a{(\dot{x})^{(n)}}\,P_a
\eeq
while the integrals are
\beq
S_1^{(n)}=Q_1^{(n)}+y\,G^{(n)} \qq\qq S_2^{(n)}=Q_2^{(n)}+y\,Q_1^{(n)}+\frac{y^2}{2}\,G^{(n)}\eeq
with
\beq
Q_1^{(n)}=\sum_{k=1}^n\,\wti{b}_k^{(n)}\,H^{n-k}\,\Pi\,P_y^{2k-1} \qq\qq   Q_2^{(n)}=\sum_{k=1}^n\,\wti{c}_k^{(n)}\,H^{n-k}\,P_y^{2k}.
\eeq

We have the relations
\beq
\lim_{A_n \to 0}\,F^{(n)}=F^{(n-1)}=F^{(n-1)}=A_{n-1}\prod_{k=1}^{n-1}(a-a_k)\qq\qq \lim_{A_n \to 0}\,G^{(n)}=P_y^2\,G^{(n-1)}.\eeq
It follows from the ODE for $x(a)$ that
\beq
x^{(n)}=\sum_{i=1}^n\,\frac{\xi_i}{\sqrt{\De_i}}\ \to\ x^{(n-1)}=\sum_{i=1}^{n-1}\,\frac{\xi_i}{\sqrt{\De_i}}\quad\Longrightarrow\quad H^{(n)}\ \to\ H^{(n-1)}.
\eeq
Hence to take the limit properly we have to let $\ A_n\to 0$ and $\xi_n\to 0$. Let us denote this limit by the symbol LIM. We will now prove:

\begin{nth} In the limit $\ A_n\to 0$ and $\xi_n\to 0$ the integrals $Q_1^{(n)}$ and $Q_2^{(n)}$ become reducible according to the relations:
\beq
{\rm LIM}\ Q_1^{(n)}=P_y^2\,Q_1^{(n-1)} \qq {\rm LIM}\ Q_2^{(n)}=P_y^2\,Q_2^{(n-1)}.
\eeq
Factoring out by $P_y^2$ the integrals, we have exhibited  the cascading between the following SI systems:
\beq\barr{l}
\{H^{(n)},\,P_y,\,Q_1^{(n)}\}\ \to\ \{H^{(n-1)},\,P_y,\,Q_1^{(n-1)}\}\\[4mm] 
\{H^{(n)},\,P_y,\,Q_2^{(n)}\}\ \to\ \{H^{(n-1)},\,P_y,\,Q_2^{(n-1)}\}.\earr
\eeq
\end{nth}

\nin{\bf Proof:} From Theorem 1 we have
\[
{\rm LIM}\ Q_1^{(n)}=\sum_{k=1}^n\,\Big({\rm LIM}\ \wti{b}_k^{(n)}\Big)\,(H^{(n-1)})^{n-k}\,P_y^{2k-1}.
\]
The first term in the sum vanishes: 
\[
\wti{b}_1^{(n)}=-A_n\sum_{i=1}^n\,\frac{\xi_i}{\sqrt{\De_i}}
\quad\Longrightarrow\quad {\rm LIM}\ 
\wti{b}_1^{(n)}=0.\]
Substituting $l=k-1$ in the remaining sum we get
\[{\rm LIM}\ Q_1^{(n)}=P_y^2\,\sum_{l=1}^{n-1}
\Big({\rm LIM}\ \wti{b}_{l+1}^{(n)}\Big)\,(H^{(n-1)})^{n-1-l}\,P_y^{2l-1}
\qq \wti{b}_{l+1}^{(n)}=(-1)^{l+1}\sum_{i=1}^n\frac{\xi_i}{\sqrt{\De_i}}\,A_n\,\si^{i (n)}_l.\]
Using relation (\ref{siik}) in Appendix B we have
\beq\label{redsigmai}
\lim_{A_n\to 0}\ A_n\,\si^{i (n)}_l=(-1)^l\sum_{s=1}^l\,a_i^{l-s}A_{n-s}=(-1)^{l-1}\sum_{t=0}^{l-1}a_i^{l-1-t}A_{n-1-t}=-A_{n-1}\,\si^{i (n-1)}_{l-1}
\eeq
which leads to
\[
{\rm LIM}\ \wti{b}_{l+1}^{(n)}=(-1)^l\sum_{i=1}^{n-1}\frac{\xi_i}{\sqrt{\De_i}}\,A_{n-1}\si^{i (n-1)}_{l-1}=\wti{b}_l^{(n-1)}\quad\Longrightarrow\quad {\rm LIM}\ Q_1^{(n)}=P_y^2\ Q_1^{(n-1)}.\]

From Theorem 1 we have also
\[
{\rm LIM}\ Q_2^{(n)}=\sum_{k=1}^n\Big({\rm LIM}\ \wti{c}_k^{(n)}\Big)\,(H^{(n-1)})^{n-k}\,P_y^{2k}.\]
The first term vanishes again:
\[
\wti{c}_1^{(n)}=\frac{A_n}{2}\left(\sum_{i=1}^n\,\frac{\xi_i}{\De_i}+\sum_{i\neq j=1}^n\,\frac{\xi_i\xi_j}{\sqrt{\De_i \De_j}}\right)\quad\Longrightarrow\quad {\rm LIM}\ \wti{c}_1^{(n)}=0.\]
Substituting $l=k-1$ in the remaining sum we get
\[{\rm LIM}\ Q_2^{(n)}=P_y^2\,\sum_{l=1}^{n-1}
\Big({\rm LIM}\ \wti{c}_{l+1}^{(n)}\Big)\,(H^{(n-1)})^{n-1-l}\,P_y^{2l-1}\]
with
\beq\label{limQ2}
\wti{c}_{l+1}^{(n)}(a)=\frac{(-1)^l}{2}\left(\sum_{i=1}^n\frac{\xi_i^2\ A_n\,\si^{i(n)}_{l}}{\De_i}+
\sum_{i\neq j=1}^n\,\frac{\xi_i\,\xi_j}{\sqrt{\De_i\,\De_j}}\Big(A_n\,\si^{ij(n)}_{l}+a\,A_n\,\si^{ij(n)}_{l-1}\Big)\right)
\eeq
Let use relation (\ref{id3sfr})  
\[
A_n\,\si^{ij(n)}_l=-\frac{A_n\,\si^{i(n)}_{l+1}-A_n\,\si^{j(n)}_{l+1}}{a_i-a_j}\]
which, combined with (\ref{redsigmai}), leads to
\[
\lim_{A_n \to 0}\,A_n\,\si^{ij(n)}_l=\frac{A_{n-1}\,\si^{i(n-1)}_l-A_{n-1}\,\si^{j(n-1)}_l}{a_i-a_j}=-A_{n-1}\,\si^{ij(n-1)}_{l-1}.\]
Plugging this last relation as well as (\ref{redsigmai}) into (\ref{limQ2}) we get
\[
{\rm LIM}\ \wti{c}_{l+1}^{(n)}=\wti{c}_l^{(n-1)}\quad\Longrightarrow\quad {\rm LIM}\ Q_2^{(n)}=P_y^2\ Q_2^{(n-1)}\]
which concludes the proof.$\quad\Box$

\vspace{3mm}
\nin{\bf Remarks:} 
\brm
\item The cascading process reduces the degree of the integrals from $2n$ to $2(n-1)$.
\item Let us start from the iff equations (\ref{condSint}) for integrals of degree $2n$
\[
\{H,Q_1^{(n)}\}+2a\,P_y\,G^{(n)}=0\qq\qq \{H,Q_2^{(n)}\}+2a\,P_y\,Q_1^{(n)}=0.\]
The cascading substitutions $Q_1^{(n)}\to P_y^2\,Q_1^{(n-1)}$ and $Q_2^{(n)}\to P_y^2\,Q_2^{(n-1)}$ in these relations and factoring by $P_y^2$ lead to iff equations for integrals of degree $2(n-1)$.
\item We have given a direct proof of the cascading phenomenon based on the explicit form of the $\wti{b}_k^{(n)}$ and $\wti{c}_k^{(n)}$. This constitutes a check of the formulas obtained above for these coefficients when $F$ is simple.
\erm

\section{Some globally defined examples}
A look at the metric
\beq\label{gmet}
g=\left(\frac{\dot{x}}{a}\right)^2\,da^2+\frac 1a\,dy^2
=\frac{dx^2}{a^2}+\frac 1a\,dy^2\eeq
shows that

\begin{nth} The metric is riemannian iff $a>0$.
\end{nth}

In general $a$ will take values in some interval $(a_m,\,a_M)$ with $a_m>0$. The end-points of this interval may be  of two kinds:
\brm
\item True singularities, namely curvature singularities, which cannot be disposed of by some coordinates change and which do prevent the metric to be defined on a manifold. They can be detected from the behaviour in a neighbourhood of these points of the scalar curvature given by (\ref{scalR}).
\item Apparent singularities, as for instance
\[g\sim d\chi^2+\chi^2\,dy^2\qq\quad \chi \to\ 0+\qq y\in{\mb S}^1\]
which can be removed using cartesian coordinates
\[ x=\chi\,\cos y\qq\qq y=\chi\,\sin y\qq 
\qq\Longrightarrow\qq g\sim dx^2+dy^2\]
\erm

Let us prove

\begin{nth}\label{sings} We have the following possibilities:
\brm
\item[a)] If $\dot{x}(a)\sim (a-a_1)^{\alf}$ the point $a=a_1\neq 0$ is a curvature singularity  if $\alf>1$. 
\item[b)] If $\dot{x}(a)\sim a^{\alf}$ then $a=0+$ is a curvature singularity if $\alf>0$.
\item[c)] If $\dot{x}(a)\sim a^{\alf}$ then $a \to +\nf$ is a curvature singularity if $\alf\in(-\nf,-1/2)\cup-1/2,0)$.
\erm
\end{nth}

\nin{\bf Proof:} In the case a) the computation of the curvature gives
\[-2R\sim \frac{2a_1\,\alf}{(a-a_1)^{2(\alf-1)}}\]
which proves the statement. 

In the case b) we have
\[
-2R\sim \frac{(2\alf+1)}{a^{2\alf}}\]
which is not continuous for $a\to 0+$ if $\alf>0$.

In the case c) the same formula holds but now 
$a\to +\nf$ and the curvature must remain bounded, which is excluded iff $\alf<0$ and $\alf\neq -1/2$. The case $\alf=-1/2$ needs a specific analysis for each metric.
$\quad\Box$ 

As a first check let us consider the simplest case $n=1$ for which the integrals are merely quadratic: we should recover one of the Koenigs metrics \cite{VaK}.

\begin{nth} For $n=1$, i. e. quadratic integrals, there is a single SI metric,  globally defined (g. d.) on ${\mb H}^2$: it is the Koenigs metric of type 3.
\end{nth}

\nin{\bf Proof:} Here we have $F(a)=a-a_1$ so we need to order the discussion according to the values taken by $a_1$. If  $a_1>0$, we may take $a_1=1$ and $\xi_1=1$. Hence we have
\[x(a)=\frac 1{\sqrt{a+1}}\qq a>0,\]
and by Proposition 13 this metric is singular for $a\to +\nf$. Indeed the coordinate change
\[
t=\sqrt{\frac a{a+1}}\qq\Longrightarrow\qq g=(1-t^2)\frac{dt^2+dy^2}{t^2}\qq t\in (0,1)\qq y\in{\mb R}\]
shows that $t\to 1-$ (i.e. $a\to +\nf$) is a curvature singularity of the metric because the conformal factor does vanish. The same argument applies if $a_1=0$. 
 
If $a_1<0$, up to scalings, we may take $a_1=-1$ and $\xi_1=1$. A first possible case is
\[x=\frac 1{\sqrt{a-1}}\qq a\in(1,+\nf)\]
and here too Proposition 13 shows that this metric is singular for $a\to +\nf$. 

The last possible case is 
\[x=\frac 1{\sqrt{1-a}}\qq\qq a\in(0,1)\qq \Longrightarrow\qq 
g=\frac 1a\left(\frac{da^2}{4a(1-a)^3}+dy^2\right).\]
The coordinate change $\dst u=\sqrt{\frac a{1-a}}$ 
shows that 
\[
g=(1+u^2)\ \frac{du^2+dy^2}{u^2}=(1+u^2)\,g_0(H^2,{\cal P})\qq u>0\qq y\in{\mb R}\]
where $g_0$ is the Poincar\'e half-plane model of ${\mb H}^2$. The resulting hamiltonian
\[ 
H=\frac{u^2}{1+u^2}\Big(P_u^2+P_y^2\Big)\]
is nothing but Koenigs metric of type 3 as it is written in Theorem 7 of \cite{VaK} (setting $\xi=0$) which was shown to be globally defined on the manifold $M\cong {\mb H}^2$. $\quad\Box$

This Koenigs metric of type 3 suggests the following generalization to SI systems with integrals of any even degree larger than 4:

\begin{nth} Let us consider, for $n\geq 2$
\beq F(a)=(a-1)\,\wh{F}(a) \qq\qq \wh{F}(a)=\prod_{i=2}^n(a-a_k)\qq\qq 0<a<1\eeq
with 
\[a_i<0\ \vee \  a_i>1 \qq i=2,\ldots n.\]
The associated SI system produces the metric
\beq\label{ex1met}
g=(1+u^2)\ \frac{\mu^2(u)\,du^2+dy^2}{u^2}\qq\qq  
u\in\,(0,+\nf)\quad y\in\,{\mb R},
\eeq
where
\beq\label{ex1f}
\mu(u)=c+\sum_{i=2}^n\frac{\xi_i}{(1+\rho_i\,
u^2)^{3/2}} 
\eeq
with 
\[
c> 0 \qq\qq\xi_i> 0\qq\qq \rho_i\in(0,1)\cup(1,+\nf).\]
This metric and the related integrals $(S_1,\,S_2)$ are globally defined on $M\cong {\mb H}^2$.
\end{nth}

\nin{\bf Proof:} The choice made for $F(a)$ and Proposition \ref{propode} imply that we may take
\beq
x(a)=\frac{c}{\sqrt{1-a}}-\sum_{i=2}^n\frac{\eps_i(-\eps_ia_i)^{3/2}\,\xi_i}{\sqrt{\eps_i(a-a_i)}}.
\eeq 
It follows that, under the coordinate change
\[ u=\sqrt{\frac a{1-a}}\qq\qq a\in\,(0,1)\ \to\ u\in\,(0,+\nf),\qq\Longrightarrow\qq \frac{dx}{\sqrt{a}}=\mu(u)du\]
where $\mu$ is given in (\ref{ex1f}). Under this substitution the metric takes the form given in the formula (\ref{ex1met}). The parameters $\rho_i$ which appear are given by
\[\rho_i=1-\frac 1{a_i}\qq\Longrightarrow\qq \rho_i\in\,(0,1)\cup(1,+\nf).\]
Defining the coordinate change
\[t=u\,\Om(u) \qq \Om(u)=c+\sum_{i=2}^n\,\frac{\xi_i}{\sqrt{1+\rho_i\,u^2}}:\qq u\in(0,+\nf)\to\ t\in(0,+\nf),\]
since $\frac{dt}{du}=\mu>0$ it follows that the inverse function $u(t)$ is increasing and $C^{\nf}([0,+\nf))$. The metric becomes
\[ g=(1+u^2(t))\Om^2(t)\ \frac{dt^2+dy^2}{t^2}=(1+u^2(t))\Om^2(t)\ g_0(H^2,{\cal P})\]
and since the conformal factor $(1+u^2(t))\Om^2(t)$ never vanishes we conclude that $M\cong {\mb H}^2$.

We have seen that the integrals are
\beq\label{Sgd}
S_1=Q_1+y\,G\qq S_2=Q_2+y\,Q_1+\frac{y^2}{2}\,G\qq G=\sum_{k=0}^n\,A_n\,H^k\,P_y^{2(n-k)}.\eeq
The global structure of these integrals is easy to study because $(H,\,P_y,\,\Pi)$, hence $G$, are globally defined on $M$. Let us consider  
\[
Q_1=\sum_{k=1}^n\wti{b}_k\,H^{n-k}\,\Pi\,P_y^{2k-1}\qq\qq 
\wti{b}_k=(-1)^k\,\left(\frac{c\ \si^1_{k-1}}{\sqrt{1-a}}-\sum_{i\in I}\frac{\eps_i(-\eps_i a_i)^{3/2}\xi_i\ \si^i_{k-1}}{\sqrt{\De_i}}\right)\]
which becomes in terms of the coordinate $t$ 
\[\forall k\in {\cal S}^1_n: \qq  
\wti{b}_k=(-1)^k\,\sqrt{1+u^2(t)}\left(c\, \si^1_{k-1}+\sum_{i\in I}\frac{a_i\, \xi_i\ \si^i_{k-1}}{\sqrt{1+\rho_i\,u^2(t)}}\right),\]
showing that all these coefficients are $C^{\nf}([0,+\nf))$ so that $Q_1$ hence $S_1$ are globally defined on $M\cong{\mb H}^2$. 

Let us consider $Q_2$. We have
\[ Q_2=\sum_{k=1}^n\wti{c}_k\,H^{n-k}\,P_y^{2k}\]
and the coefficients become
\[\barr{l}\dst \forall k\in {\cal S}^1_n: \qq  \wti{c}_k=\frac{(-1)^{k-1}}{2}(1+u^2(t))\left(c^2\si^1_{k-1}+2c\sum_{i\in I}\frac{a_i\,\xi_i}{\sqrt{1+\rho_i\,u^2(t)}}\tau^{1i}_{k-1}+\right.
\\[5mm]\dst 
\hspace{4cm}\left.+\sum_{i\in I}\frac{a_i^2\,\xi_i^2}{1+\rho_i\,u^2(t)}\si^i_{k-1}+\sum_{i\neq j\in I}\frac{a_i\,a_j\,\xi_i\,\xi_j}{\sqrt{(1+\rho_i\,u^2(t))(1+\rho_j\,u^2(t))}}\tau^{ij}_{k-1}\right)\earr\]
with
\[\tau^{ij}_{k-1}=\si^{ij}_{k-1}+\frac{u^2}{1+u^2}\,\si^{ij}_{k-2}.\]
From this formula it follows that $Q_2$ hence $S_2$ are globally defined on $M\cong {\mb H}^2$.$\quad\Box$

As a second example we have:

\begin{nth}\label{R2} For $n\geq 2$ the choice
\[
F(a)=(a-a_1)(a-a_2)\wh{F}(a) \qq\qq 0<a_1<a<a_2\]
with \footnote{If $n=2$ we have $\wh{F}(a)=1$.}
\[\wh{F}(a)=\prod_{i=3}^n(a-a_i): \qq \qq \Big(\ a_i<a_1\ \vee \  a_i>a_2 \qq i=3,\ldots n\ \Big)\]
leads to a SI system with the metric
\beq\label{ex2met}
g=\frac 1{A(t)}(dt^2+dy^2)\qq\qq (t,y)\in{\mb R}^2
\eeq
globally defined on the manifold $M\cong{\mb R}^2$ as well as the integrals $S_1$ and $S_2$.
\end{nth}

\nin{\bf Proof:} Let us consider \footnote{For $n=2$ the last sum is absent.}
\[
x(a)=-\frac{\wti{\xi}_1}{\sqrt{a-a_1}}+\frac{\wti{\xi}_2}{\sqrt{a_2-a}}-\sum_{i=3}^n\frac{\eps_i\,\wti{\xi}_i}{\sqrt{\eps_i(a-a_i)}}\]
with
\[
\wti{\xi}_1=a_1\sqrt{a_2-a_1}\,\xi_1\qq\qq \wti{\xi}_2=a_2\sqrt{a_2+a_1}\,\xi_2 \qq \qq \wti{\xi}_i=a_i\sqrt{a_2-a_1}\,\xi_i\]
and
\[
\eps_i=+1\quad a_i<a_1\qq \& \qq \eps_i=-1 \quad a_i>a_2.\]
The coordinate change
\[
a=a_1+(a_2-a_1)s^2 \qq s\equiv \sin\tht: \quad a\in(a_1,a_2)\,\leftrightarrow\,\tht\in (0,\pi/2)\]
gives
\[
x(\tht)=-\frac{\xi_1\,a_1}{s}+\frac{\xi_2\,a_2}{\sqrt{1-s^2}}-\sum_{i=3}^n\,\frac{\eps_i a_i\xi_i}{\sqrt{\eps_i(\rho_i+s^2)}} \qq\qq \rho_i=\frac{a_1-a_i}{a_2-a_1}.\]
So differentiating we get
\[
D_{\tht}\,x=\xi_1\frac{a_1\,c}{s^2}+\xi_2\frac{a_2\,s}{(1-s^2)}+\sum_{i=3}^n\,\frac{a_i\xi_i\, s\, c}{(\rho_i+s^2)^{3/2}}\]
These relations show that $D_{\tht}\,x>0$.

Defining $\dst dt=\frac{dx}{\sqrt{a}}$ one gets
\[
t(s)=\sqrt{a_1+(a_2-a_1)s^2}\left(-\frac{\xi_1}{s}+\frac{\xi_2}{\sqrt{1-s^2}}-\sum_{i=3}^n\,\frac{\eps_i\xi_i}{\sqrt{\rho_i+s^2}}\right).\]
It follows that $\tht\in(0,\pi/2)$ is mapped into $t\in{\mb R}$ and that the function $t=h(s)$ is a $C^{\nf}$ (increasing) bijection from $\tht\in(0,\pi/2)\to {\mb R}$, hence its inverse function $s=h^{-1}(t)$ is also a $C^{\nf}$ (increasing) bijection.

We obtain the metric given by (\ref{ex2met}):
\[
g=\frac 1{A(t)}(dt^2+dy^2)\qq\qq A=a\circ h^{-1}\qq\qq 
(t, y)\in{\mb R}^2\]
and since the conformal factor $A(t)$ never vanishes, we  conclude that the manifold is $M\cong {\mb R}^2$.

As in the previous case considered above, let us consider  
\[
Q_1=\sum_{k=1}^n\wti{b}_k\,H^{n-k}\,\Pi\,P_y^{2k-1}\]
with the coefficients
\[
\wti{b}_k=(-1)^k\,\left(-\frac{a_1\,\xi_1\,\si^1_{k-1}}{A(t)}+\frac{a_2\,\xi_2\,\si^2_{k-1}}{\sqrt{1-A^2(t)}}-\sum_{i=3}^n\frac{\eps_i\xi_i\,\si^i_{k-1}}{\sqrt{\eps_i(\rho_i+A^2(t))}} \right).\]
Since $\tht\in(0,\pi/2)$ this proves that $Q_1$ is indeed globally defined. The check for $Q_2$ is similar.
$\quad\Box$

Let us conclude with the following negative result:

\begin{nth}\label{nogo} The choice $F=(a^2+a_0^2)^n$ for $n=1,2,\ldots$ never leads to a globally defined SI system.
\end{nth}

\nin{\bf Proof:}  In this case, setting $a_0=1$, we have
\[
x(\tht)=\sum_{k=1}^n\Big(\mu^+_k(\cos\tht)^{k-1/2}\,\cos((k-3/2)\tht)+\mu^-_k(\cos\tht)^{k-1/2}\,\sin((k-3/2)\tht)\Big).\]
with $a=\tan\tht$ and $\tht\in\,(0,\frac{\pi}{2})$. Let us consider the metric behavior for $\tht\to \frac{\pi}{2}-$ for a fixed value of $k$ which can always be obtained by an appropriate choice of the coefficients $\mu^{\pm}_k$. We have the following equivalent:
\[
g\sim c_k^2(\cos\tht)^{2k-1}\,d\tht^2+\cos\tht\,dy^2.\]
Using the coordinate change $v=v_0(1-\sin\tht)^{(2k+1)/4}$ for an appropriate constant $v_0$ leads to
\[
g\sim c_k^2\Big(dv^2+v^{2/(2k+1)}\,dY^2\Big)\qq\quad v\to 0+\qq\quad k\in\{1,2,\ldots,n\}\]
where $Y$ is merely homothetic to $y$. Even restricting $Y\in\,{\mb S}^1$ and no matter one chooses $k$ the exponent of $v$ will never be equal to $2$ so we have a true singularity for $v\to 0+$.$\quad\Box$

\newpage 
\part{Integrals of odd degree in the momenta}

\vspace{4mm} 
We will consider the cases for which the observables have for degree in the momenta $\sharp(Q)=2n+1$ 
where $n\geq 1$. The case $n=1$ was first analyzed in \cite{ms} and \cite{vds}.

The hamiltonian remains unchanged
\beq
H=\Pi^2+a\,P_y^2 \qq\qq \Pi=\frac a{\dot{x}}\,P_a,\eeq
while 
\beq
G=\sum_{k=0}^n\,A_k\,H^k\,P_y^{2(n-k)+1}\qq A_n\neq 0\qq\qq\qq \sharp(G)=2n+1\qq n\in{\cal S}_n\eeq
is still built up from the $n$ constants: $\ A_0,\,A_1,\cdots,A_{n-1}$. The SI stems from the two observables 
\beq
S_1=Q_1+y\,G \qq\qq S_2=Q_2+y\,Q_1+\frac{y^2}{2}\,G.
\eeq
Let us begin with the determination of $Q_1$.

\section{The local structure of the first integral}

\begin{nth} Taking  
\beq
Q_1=\sum_{k=0}^n\,b_k(a)\,\Pi^{2k+1}\,P_y^{2(n-k)},\eeq
the observable $S_1$ will be an integral iff
\beq\label{sysO}\left\{\barr{ll} (a) & \dst\qq b_0=2F\ \dot{x}\\[4mm] 
(b) & \qq\dst \dot{b}_{k-1}=(k+\frac 12)\,b_k-\frac{D_a^k\,F}{k!}\ \dot{x}\qq k=1,\ldots,n\\[4mm]
(c) & \qq\dst \dot{b}_n=0\earr\right.\eeq
where $\dst \ F(a)=\sum_{k=0}^n\,A_k\,a^k.$
\end{nth}

\nin{\bf Proof:} The equation 
\[\{H,S_1\}=\{H,Q_1\}+2a\,P_y\,G=0\]
expands into
\[\sum_{k=1}^{n+1}\,a\,b'_{k-1}\Pi^{2k}\,P_y^{2(n-k+1)}-\sum_{k=0}^n(k+1/2)a'b_k\,a\,\Pi^{2k}\,P_y^{2(n-k+1)}
+\sum_{k=0}^n\frac{D_a^k\,F}{k!}\,a\,\Pi^{2k}\,P_y^{2(n-k+1)}=0\]
giving the differential system
\[\left\{\barr{l}\dst 0=\frac 12\,a'\,b_0-F(a)\\[4mm]\dst  
b'_{k-1}=(k+\frac 12)\,a'\,b_k-\frac{D_a^k\,F}{k!}\qq k=1,\ldots,n \\[4mm]\dst 
b'_n=0\earr\right.\]
Switching to the new variable $a$, instead of $x$, gives (\ref{sysO}).$\quad\Box$
 
We can proceed to

\begin{nth}[Linearizing ODE]\label{propode2} The differential system (\ref{sysO}) has for (unique) solution
\beq
k\in{\cal S}_n^0\bs\{n\}: \qq b_k=
\sum_{s=1}^{k+1}\frac{D_a^{k-s+1}F}{(k-s+1)!}\,\frac{D^s_a\,x} {(1/2)_s}\qq\qq 
b_n={\rm const}=\nu_n\in{\mb R}\bs \{0\}
\eeq
where $x(a)$ is a solution of the ODE
\beq\label{solO}
{\rm Op}_n[F]\,x(a)=
\left(n+\frac 12\right)\nu_n\,a+\be_n\qq\qq \be_n\in{\mb R}.
\eeq
Its solution in the simple case, up to an additive constant, is given by
\beq
x=\frac{\nu_n}{2}\,a+\sum_{i=1}^n\frac{\xi_i}{\sqrt{\De_i}} \qq \De_i=\eps_i(a-a_i)\qq \eps_i^2=1\qq \xi_i\in{\mb R}.\eeq
\end{nth}

\nin{\bf Proof:} The recursive proof giving $b_k$ for any $k\in{\cal S}^0_n$ is similar to the one given 
for Proposition \ref{Q1bk}. Integrating equation (b) in (\ref{sysO}) for $k=n$ leads to
\[b_{n-1}+\frac{D_a^n\,F}{n!}\,x=(n+1/2)\nu_n\,a+\be_n\]
which we combine with
\[b_{n-1}=\sum_{s=1}^n\frac 1{(1/2)_s}\frac{D_a^{(n-s)}\,F}{(n-s)!}\,D_a^s\,x\]
to get (\ref{solO}). The homogeneous equation was already solved for in Proposition \ref{propode} and looking for 
an affine solution gives 
\[\frac{\nu_n}{2}\,a+\be_n-\nu_n\,A_{n-1}\]
in which the constant term may be deleted. $\quad\Box$

\vspace{3mm}
\nin{\bf Remarks:} 
\brm
\item The transition from integrals of degree $2n$ to $2n+1$ is strikingly simple: one just adds a linear term in  $x$! This was observed in \cite{vds} for the cubic case but was not expected to be so general.
\item In Proposition \ref{ccurv} we have seen that if $\dst x=\pm\frac a{\sqrt{-2R}}$ the metric is of constant negative curvature. In order to avoid such a case we must exclude the trivial possibility that all the $\xi_i$ be vanishing. 
\erm
Let us conclude this section by giving a useful form of $Q_1$:

\begin{nth}\label{bkt92} For a generic choice of $F$  one can write
\beq\label{finQ1}
Q_1=\sum_{k=0}^n\ \wti{b}_k[F]\,\Pi\,H^{n-k}\,P_y^{2k}\eeq
with
\beq\label{detailQ1}
\forall k \in\ {\cal S}^0_n: \qq\wti{b}_k[F]=\sum_{s=0}^k\,{n-s \choose k-s}(-a)^{n-k}\,b_{n-s}[F].\eeq
For a simple $F$ we have
\beq\label{bkt22}
\forall k\in{\cal S}_n^0: \qq\qq \wti{b}_k[\wh{F}]=(-1)^k\,\left(\nu_n\,\si_k+\sum_{i=1}^n\frac{\xi_i}{\sqrt{\De_i}}\,\si^i_{k-1}\right).
\eeq
\end{nth}

\nin{\bf Proof:} By the same argument used in Part I for $Q_1$ one gets relation (\ref{detailQ1}). For a simple 
$F$ we have
\[\wti{b}_k[\wh{F}]=\sum_{s=0}^k\,{n-s \choose n-k}(-a)^{k-s}\,b_{n-s}[\wh{F}]={n \choose n-k}(-a)^k\,\nu_n+\sum_{s=1}^k{n-s \choose n-k}(-a)^{k-s}\,b_{n-s}[\wh{F}]\]
where
\[b_{n-s}[\wh{F}]=\nu_n\,\frac{D_a^{n-s}\,\wh{F}}{(n-s)!}
+\sum_{l=1}^{n-s+1}\frac 1{(1/2)_l}\frac{D_a^{n-s+1-l}\,\wh{F}}{(n-s+1-l)!}\,D_a^l\,x^{(0)}\]
where $x^{(0)}$ is just the $\xi_i$ dependent part of $x$. So we have to compute two pieces:
\[\wti{b}_k[\wh{F}]=\nu_n\sum_{s=0}^k\,{n-s \choose n-k}(-a)^{k-s}\,\frac{D_a^{n-s}\,F}{(n-s)!}
+\sum_{s=0}^k\,{n-s \choose n-k}(-a)^{k-s}\,b_{n-s+1}({\rm even})\]
where $\,b_{n-s+1}({\rm even})$ is the same as in the proof of the Proposition \ref{Q1bk} and therefore gives 
the same result as in Proposition \ref{bt9}:
\[(-1)^k\,\sum_{i=1}^n\frac{\xi_i\,\si^i_{k-1}}{\sqrt{\De_i}}.\]
The first piece gives
\[\nu_n\sum_{s=0}^k\sum_{l=n-s}^n\,A_l\,{n-s \choose n-k}(-a)^{k-s}\,{l \choose n-s}
(-1)^{k-s}\,a^{k+l-n}.\]
Reversing the summations we conclude to
\[\nu_n\sum_{l=n-k}^n\,A_l\,a^{k+l-n}\frac{l!}{(n-k)!}\,
\sum_{s=n-l}^k\,\frac{(-1)^{s-k}}{(k-s)!\,(l+s-n)!}=\nu_n\,A_{n-k}=(-1)^k\,\nu_n\,\si_k\,\]
and use of the identity (\ref{sik}) concludes the proof.
$\quad\Box$

As in Part I, let us check the result obtained for $\wti{b}_k[\wh{F}]$ using its differential system. We have

\begin{nth}Defining
\beq
\forall k \in\,{\cal S}^0_n: \qq 
\wti{b}_k[F]=(-1)^k\,\be_k[F],\eeq 
for any choice of $F$ the relations 
\beq\label{sysbe2}\barr{ll}
  & \qq \dot{\be}_0=0\\[4mm]\dst 
0\leq k\leq n-1: & \qq\dot{\be}_{k+1}=-a\,\dot{\be}_k-\frac 12\,\be_k+\si_k\,\dot{x}\\[4mm] 
k=n:  & \qq 0=-a\,\dot{\be}_n-\frac 12\,\be_n+\si_n\,\dot{x}\earr\eeq
imply that $S_1$ is an integral. For a simple $F$ the formula obtained for $\wti{b}_k[\wh{F}]$ in Proposition \ref{bkt92} is indeed a solution of this system. 
\end{nth}

\vspace{3mm}
\nin{\bf Proof:} a routine computation leads to
\[\barr{l}\dst 
0=\{H,Q_1\}+2a\,P_y\,G=\frac a{\dot{x}}\,(2\dot{\be}_0)\,H^{n+1}-\frac a{\dot{x}}\,\sum_{k=0}^{n-1}(-1)^k(2\dot{\be}_{k+1})\,H^{n-k}\,P_y^{2(k+1)}\\[4mm]\dst \hspace{6cm} 
+\frac a{\dot{x}}\,\sum_{k=0}^n(-1)^k \Big(-\be_k-2a\dot{\be}_k+2(-1)^kA_{n-k}\,\dot{x}\Big)H^{n-k}\,P_y^{2(k+1)}\earr\]
from which we deduce (\ref{sysbe2}) using the identity (\ref{sik}).

Let us check that for a simple $F$ the relation (\ref{bkt22}) does give a solution of this differential system. We have first $\be_0=\nu_n$ which is fine. Then computing
\[
-a\dot{\be}_k-\frac 12\,\be_k+\si_k\,\dot{x}=\frac 12\sum_{i=1}^n\,\frac{\eps_i\,\xi_i}{\De_i^{3/2}}\,(a_i \si^i_{k-1}-\si_k)=-\frac 12\sum_{i=1}^n\,\frac{\eps_i\,\xi_i}{\De_i^{3/2}}\,\si^i_k \]
So for $0\leq k \leq n-1$ we get $\dot{\be}_{k+1}$ while for $k=n$ it indeed vanishes.$\quad\Box$

In fact there is a simple structural relation between the $\Big(x^{\neq},\ \be^{\neq}_k[F]\Big)$ with $k=0,1,\ldots,n$ for odd degree integrals and $\Big(x^{=},\ \be^{=}_k[F]\Big)$ with $k=1,\ldots,n$ for even degree integrals given by:

\begin{nth} The relations
\beq\label{oddeven}
\left\{\barr{l}
x^{\neq}(a)=\frac 12\,\nu\,a+x^{=}(a)\\[4mm] 
\be^{\neq}_0[F]=\nu\qq\qq
\forall k\in\,{\cal S}_n^1:\qq \be^{\neq}_k[F]=\nu\,\si_k+\be^{=}_k[F]\earr\right.
\eeq
ensure that the equations (\ref{sysbe2}) for the $\be^{\neq}_k[F]$ imply the equations (\ref{sysbe}) for the $\be^{=}_k[F]$.
\end{nth}

\nin{\bf Proof:} The first relation in (\ref{oddeven}) follows from Propositions \ref{propode} and \ref{propode2}.
For $k=0$ we have  
\[\dot{\be}^{\neq}_1[F]=\dot{\be}^{=}_1[F]=-\frac 12\,\nu+\si_0\Big(\frac{\nu}{2}+\dot{x}^{=}\Big)=\dot{x}^{=}.\]
For $1\leq k\leq n-1$ we have 
\[\dot{\be}^{\neq}_{k+1}[F]=-a\,\dot{\be}^{\neq}_k[F]-\frac 12\,\be^{\neq}[F]+\si_k\,\dot{x}^{\neq}=-a\,\dot{\be}^{=}_k[F]-\frac 12\,\Big(\nu\,\si_k+\be^{=}_k[F]\Big)+\si_k\,\Big(\frac{\nu}{2}+\dot{x}^{=}\Big)\]
which is indeed equal to $\dot{\be}^{=}_{k+1}[F]$. The argument for $k=n$ is similar.$\quad\Box$

Now that $\,Q_1$ is fixed up let us construct $\,Q_2$.

\section{The local structure of the second integral}
As shown in Proposition 1, the structure of $Q_2$ follows from:
\beq
\{H,Q_2\}+2a\,P_y\,Q_1=0 \qq\quad Q_2=\sum_{k=0}^n\wti{c}_k[F]\,H^{n-k}\,P_y^{2k+1}\eeq
and it is given by:

\begin{nth} The observable $S_2$ is an integral iff $Q_2$ is determined from the differential system
\beq\label{eqintQ2}
\forall k\in{\cal S}_n^0:\qq\qq D_a\,\wti{c}_k=-\wti{b}_k\,\dot{x}.\eeq
For a simple $F$ these coefficients are given by
\beq\barr{l}\dst 
\forall k\in{\cal S}_n^0:\quad\wti{c}_k[\wh{F}]=\frac{(-1)^{k+1}}{2}\left\{\nu_n^2\,a\,\si_k
+2\nu_n\sum_{i=1}^n\frac{\xi_i}{\sqrt{\De_i}}(\si^i_k+a\,\si^i_{k-1})+\right.\\[4mm]\dst 
\hspace{5.5cm}+\left.\sum_{i=1}^n\frac{\xi_i^2
}{\De_i}\,\si^i_{k-1}+
\sum_{i\neq j=1}^n\,\frac{\xi_i\,\xi_j}{\sqrt{\De_i\,\De_j}}
\Big(\si^{ij}_{k-1}+a\,\si^{ij}_{k-2}\Big)\right\}.\earr\eeq
\end{nth}

\nin{\bf Proof:} An elementary computation gives
\[\{H,Q_2\}+2a\,P_y\,Q_1=2a\Pi\sum_{k=0}^n\left(\frac{D_a\,\wti{c}_k}{\dot{x}}+\wti{b}_k\right)
H^{n-k}\,P_y^{2k+1}\]
which implies (\ref{eqintQ2}).

In the computation of $\,D_a\,\wti{c}_k$ there appears the constant term
\[(-1)^{k+1}\frac{\nu_n^2}{2}\,\si_k\]
while the terms linear in the $\xi_i$ give
\[(-1)^{k+1}\nu_n\,\left(\frac 12\sum_{i=1}^n\frac{\eps_i\,\xi_i}{\De_i^{3/2}}
+\sum_{i=1}^n\frac{\xi_i\,\si^i_{k-1}}{2\sqrt{\De_i}}\right)\]
and an integration yields
\[(-1)^{k+1}\nu_n\,\sum_{i=1}^n\frac{\xi_i}{\sqrt{\De_i}}(\si_k+(a-a_i)\si^i_{k-1})
=(-1)^{k+1}\nu_n\,\sum_{i=1}^n\frac{\xi_i}{\sqrt{\De_i}}(\si^i_k+a\,\si^i_{k-1})\]
after use of (\ref{id1sfr}). The remaining terms are quadratic in the momenta and are easily seen to be the same as in Proposition \ref{propQ2}. $\quad\Box$

Let us add an important algebraic relation:

\begin{nth} The integrals $S_1$ and $S_2$ are algebraically related by
\beq\label{idQ2}
S_1^2-2G\,S_2=\sum_{k,l=1}^n\,{\cal Q}_{kl}\,H^{2n-k-l}\,P_y^{2(k+l+1)}
+\nu_n ^2\sum_{k,l=0}^n(-1)^{k+l}\si_k\,\si_l\,H^{2n+1-k-l}P_y^{2(k+l)}\eeq
where ${\cal Q}_{kl}$ was already defined in (\ref{defqkl}).
\end{nth}

\nin{\bf Proof:} Denoting the quantities defined in the Part 1 by a sharp subscript, we have
\[G=P_y\,G_{\sharp}\qq Q_1=A+P_y\,Q_{1\,\sharp}\qq Q_2=B+P_y\,Q_{2\,\sharp}\]
which implies
\[
X\equiv S_1^2-2G\,S_2=Q_1^2-2G\,Q_2=A^2+2AP_y\,Q_{1\,\sharp}-2BP_y\,G_{\sharp}+P_y^2\,X_{\sharp}.
\]
The first three terms, after several algebraic simplifications give the required formula while the last term is obvious. $\quad\Box$

Summarizing the results obtained up to now we have
\begin{nTh} For a simple $F$, the hamiltonian
\beq
H=\Pi^2+a\,P_y^2\qq\qq \Pi=\frac{a}{\dot{x}}\,P_a\qq\qq a>0\eeq
where
\beq
x=\frac{\nu_n}{2}\,a+\sum_{i=1}^n\,\frac{\xi_i}{\sqrt{\De_i}}\qq\qq \De_i=\eps_i(a-a_i)\qq \eps_i^2=1,\eeq
with all the $\,\xi_i\in{\mb R}$ and $\,A_n\in{\mb R}\bs\{0\}$, exhibits two integrals
\beq
S_1=Q_1+y\,G \qq\qq S_2=Q_2+y\,Q_1+\frac{y^2}{2}\,G\eeq
where
\beq
G=\sum_{k=0}^n\,A_{n-k}\,H^{n-k}\,P_y^{2k+1}\quad Q_1=\sum_{k=0}^n\,\wti{b}_k\,H^{n-k}\,\Pi\,P_y^{2k} 
\quad   Q_2=\sum_{k=0}^n\,\wti{c}_k\,H^{n-k}\,P_y^{2k+1}
\eeq
with
\beq
\forall k \in {\cal S}^0_n:\quad  
\wti{b}_k[\wh{F}]=(-1)^k\,\left(\nu_n\,\si_k+\sum_{i=1}^n\,\frac{\xi_i\ \si^i_{k-1}}{\sqrt{\De_i}}\right)
\eeq
and
\beq
\barr{l}\dst 
\forall k\in{\cal S}_n^0:\quad\wti{c}_k[\wh{F}]=\frac{(-1)^{k+1}}{2}\left\{\nu_n^2\,a\,\si_k
+2\nu_n\sum_{i=1}^n\frac{\xi_i}{\sqrt{\De_i}}(\si^i_k+a\,\si^i_{k-1})+\right.\\[4mm]\dst 
\hspace{5.6cm}+\left.\sum_{i=1}^n\frac{\xi_i^2\ \si^i_{k-1}}{\De_i}+
\sum_{i\neq j=1}^n\,\frac{\xi_i\,\xi_j}{\sqrt{\De_i\,\De_j}}
\Big(\si^{ij}_{k-1}+a\,\si^{ij}_{k-2}\Big)\right\}.\earr
\eeq
These two integrals generate  two possible (maximally) SI systems:
\beq
{\cal I}_1=\{H,\,P_y,\,S_1\}\qq\qq \mbox{and}\qq\qq {\cal I}_2=\{H,\,P_y,\,S_2\}.
\eeq
\end{nTh}

\nin{\bf Proof:} The functional independence proof is the same as for Theorem 1.$\quad\Box$

\section{Some globally defined examples} 
Let us begin with the case $n=1$ (cubic integrals) for which the global structure was first analyzed in \cite{vds}. To compare most  conveniently with our results let us first transform our metric
\beq
g=\frac{\dot{x}^2}{a^2}\,da^2+\frac{dy^2}{a}\qq\quad a>0\eeq
under the coordinate change $u=\sqrt{a}$. We get
\beq\label{genmet}
g
=\frac{\mu^2\,du^2+dy^2}{u^2}\qq\qq \mu=2\dot{x}(a=u^2).
\eeq
So, for $F(a)=a-a_1$ we may take
\beq
x(a)=\frac{\nu}{2}\,a+\frac{\eps\,c}{\sqrt{\eps(a-a_1)}}
\quad\Longrightarrow\quad \mu=\nu-\frac{c}{(\eps(u^2-a_1))^{3/2}}\qq \eps=\pm 1\qq \nu\neq 0\eeq
and since $\nu$ cannot vanish we can set $\nu=1$ showing that our local form of the metric is in perfect agreement with the local form given in \cite{vds}. It may be noticed that the single difference in the function $\mu$ with respect to the Koenigs case (quadratic integrals) is just this constant $\nu$.

However, as opposed to the Koenigs metrics, this form of $\mu$ allows for a much larger number of globally defined  cases. Let us prove:

\begin{nth} The SI systems having the metric
\beq
g=\frac{\mu^2(u)\,du^2+dy^2}{u^2}\qq\qq \mu(u)=1-\frac c{[\eps(u^2-a_1)]^{3/2}}\eeq
are globally defined on $M\cong {\mb H}^2$ in the following cases:

\beq\barr{lcl}
{\cal I}_{++}: & \qq\dst \mu(u)=1-\frac 1{(u^2-a_1)^{3/2}} & \qq a_1\in\ (-\nf, -1)\\[4mm] 
{\cal I}_{+-}: & \qq\dst \mu(u)=1+\frac 1{(u^2-a_1)^{3/2}} & \qq a_1\in\ (-\nf, 0)\\[4mm] 
{\cal I}_{-+}: & \qq\dst \mu(u)=1-\frac 1{(a_1-u^2)^{3/2}} & \qq a_1\in\ (0,1)\\[4mm]
{\cal I}_{--}: & \qq\dst \mu(u)=1-\frac c{(u^2-a_1)^{3/2}} & \qq a_1\in\ (0,+\nf)\earr
\eeq

\end{nth}

\nin{\bf Proof:} The scalar curvature being
\[ R_g=-2\ \frac{\mu+u\,\mu'}{\mu^3}\]
any singularity of it implies that the metric cannot be defined on any manifold.

In the case ${\cal I}_{++}$, for $a_1\geq 0$, $\mu$ vanishes for $u_0=\sqrt{a_1+1}>\sqrt{a_1}$ leading to a curvature singularity. For $a<0$ we have $u>0$ and for 
$a_1\geq -1$ the curvature is again singular for $u_0=\sqrt{a_1+1}$. It remains to consider $a_1<-1$. Defining
\[ dt=\mu\,du \qq \Longrightarrow \qq t=u\,
\Om(u)\qq \Om(u)=1-\frac 1{|a_1|\sqrt{u^2+|a_1|}}.\]
Since $\mu=\frac{dt}{du}$ never vanishes the inverse function $u(t)$ is $C^{\nf}([0,+\nf))$. The metric is now
\[G=\Om^2(u(t))\frac{dt^2+dy^2}{t^2}=\Om^2(u(t))\,g_0(H^2,{\cal P})\]
and since $\Om([0,+\nf))=[1-1/|a_1|^{3/2},1)$, the conformal factor never vanishes showing that $M\cong{\mb H}^2$.

In the case ${\cal I}_{+-}$, for $a_1>0$ we have 
$u>\sqrt{a_1}$ so that defining
\[dt=\mu\,du \qq \Longrightarrow \qq t=u\,
\Om(u)\qq \Om(u)=1-\frac 1{a_1\sqrt{u^2-a_1}}\]
but this time $\Om(u)$  vanishes for $u_0=\sqrt{a_1+1/a_1^2}>\sqrt{a_1}$. Since $\Om(u(t))$ appears as a conformal factor in the metric there can be no manifold.  

\nin For $a_1=0$ we have $u\in\,(0,+\nf)$. In this last case, defining $v=1/3u^3$ the metric is 
\[G\sim\frac{du^2}{u^8}+\frac{dy^2}{u^2}=dv^2+(3v)^{2/3}\,dy^2\]
showing that $u \to 0+$ precludes any manifold.  

\nin For $a_1<0$, hence $u>0$, the change of coordinate
\[dt=\mu\,du \qq \Longrightarrow \qq t=u\,
\Om(u)\qq \Om(u)=1+\frac 1{|a_1|\sqrt{u^2+|a_1|}}\]  
implies for the metric
\[G=\Om^2(u(t))\frac{dt^2+dy^2}{t^2}=\Om^2(u(t))\,g_0(H^2,{\cal P}),\]
where $u(t)$ is $C^{\nf}([0,+\nf))$ and $\Om([0,+\nf))=(1,1+1/|a_1|^{3/2}]$ hence the manifold is again ${\mb H}^2$.

For $\eps=-1$ we must have $a_1>0$ and $u\in(0,\sqrt{a_1})$. 

In the case ${\cal I}_{-+}$, for $a_1\geq 1$, there is a curvature singularity for $u_0=\sqrt{a_1-1}$ while for $0<a_1<1$ the function $\mu(u)$ never vanishes. Let us define
\[dt=\mu\,du \qq \Longrightarrow \qq t=u\,
\Om(u)\qq \Om(u)=1-\frac 1{a_1\sqrt{a_1-u^2}}.\]
The function $\Om(u)$ is strictly decreasing with 
$\Om([0,\sqrt{a_1}))=(-\nf, -(1/a_1^{3/2}-1)]$ and the  
metric becomes
\[G=\Om^2(u(t))\frac{dt^2+dy^2}{t^2}=\Om^2(u(t))\,g_0(H^2,{\cal P}).\]
Since $\Om(u(t))$ never vanishes we get $M={\mb H}^2$.

In the last case ${\cal I}_{--}$, we can define
\[dt=\mu\,du \qq \Longrightarrow \qq t=u\,
\Om(u)\qq \Om(u)=1+\frac 1{a_1\sqrt{a_1-u^2}}\]
where $\Om(u)$ is strictly increasing with 
$\Om([0,\sqrt{a_1}))=[1+1/a_1^{3/2},+\nf)$ and the  
metric becomes
\[G=\Om^2(u(t))\frac{dt^2+dy^2}{t^2}=\Om^2(u(t))\,g_0(H^2,{\cal P}).\]
Since $\Om(u(t))$ never vanishes we get $M={\mb H}^2$.

For the proof that the integrals are also globally defined on ${\mb H}^2$ the arguments presented in \cite{vds} do apply and need not be repeated.$\quad\Box$

\vspace{3mm}\nin {\bf Remark:} Our analysis does correct the Propositions 16 and 17 of the reference \cite{vds} but is in agreement with its Proposition 18 . 

Let us generalize the previous SI systems:

\begin{nth}The SI systems ${\cal I}_{+\pm}$, corresponding to $F(a)=(a-a_1)$, equipped with cubic integrals, do generalize to $\dst F(a)=\prod_{i=1}^n(a-a_i)$ with integrals of degree $\ 2n+1\ $ (with $n\geq 2)$) which remains globally defined on $M\cong{\mb H}^2$ under the following restrictions
\beq\label{constr+}
\barr{llll}
{\cal I}_{++}: \quad & -\nf <a_i<a_1<-1\quad \& \qq   \xi_i>0 \qq \& \qq\dst \frac 1{|a_1|^{3/2}}+\sum_{i=2}^n\frac{\xi_i}{|a_i|^{3/2}}<1,\\[4mm]
{\cal I}_{+-}: \quad & -\nf <a_i<a_1<0\qq \& \qq \xi_i>0. \qq \earr\eeq
\end{nth}

\nin{\bf Proof:} The first SI system, above ${\cal I}_{++}$,  is generated by
\[
x(a)=\frac a2-\frac 1{\sqrt{a-a_1}}+\sum_{i=2}^n\frac{\xi_i}{\sqrt{a-a_i}}\quad\Longrightarrow\quad  \mu(u)=1-\frac 1{(u^2-a_1)^{3/2}}-\sum_{i=2}^n\frac{\xi_i}{(u^2-a_i)^{3/2}}.\]
The metric has the form (\ref{genmet}). Noting that $\mu$  
is strictly increasing from
\[
\mu(0)=1-\frac 1{|a_1|^{3/2}}-\sum_{i=2}^n\frac{\xi_i}{|a_i|^{3/2}}>0 \qq\longrightarrow\qq \mu(+\nf)=1.\]
Let us define
\[
dt=\mu(u)\,du\quad\Longrightarrow\quad t=u\,\Om(u),\qq \Om(u)=1-\frac 1{|a_1|\sqrt{u^2-a_1}}-\sum_{i=2}^n\frac{\xi_i}{|a_i|\sqrt{u^2-a_i}}.\]
Since $D_u t>0$ the inverse function $u(t)$ is $C^{\nf}([0,+\nf))$ and $\Om(u)$ is strictly increasing with $\Om([0,+\nf))=[\mu(0),1)$ and therefore never vanishes. The metric becomes
\[
g=\Om^2(u(t))\,\frac{dt^2+dy^2}{t^2}=\Om^2(u(t))\,g_0(H^2,{\cal P})\]
showing that $M\cong {\mb H}^2$.

Here we have
\[ Q_1=\sum_{k=0}^n\,\wti{b}_k\,H^{n-k}\,\Pi\,P_y^{2k}\]
with
\[\wti{b}_k=(-1)^k\,\left(\frac12\,\si_k-\frac{\si^1_{k-1}}{\sqrt{u^2(t)-a_1}}+\sum_{i=2}^n\frac{\xi_i\,\si^i_{k-1}}{\sqrt{u^2(t)-a_i}}\right)\]
which are indeed $C^{\nf}([0,+\nf))$ hence $Q_1$ is globally defined on $M$. The check for $Q_2$ is similar.

The second SI system, above ${\cal I}_{+-}$, is generated by
\[
x(a)=\frac a2-\frac 1{\sqrt{a-a_1}}-\sum_{i=2}^n\frac{\xi_i}{\sqrt{a-a_i}}\quad\Longrightarrow\quad  \mu(u)=1+\frac 1{(u^2-a_1)^{3/2}}+\sum_{i=2}^n\frac{\xi_i}{(u^2-a_i)^{3/2}}.\]
The function $\mu$ is decreasing from 
\[
\mu(0)=1+\frac 1{|a_1|^{3/2}}+\sum_{i=2}^n\frac{\xi_i}{|a_i|^{3/2} }\qq\longrightarrow\qq  \mu(+\nf)=1\]
hence it never vanishes. So we can define
\[
t=u\,\Om(u)\qq\qq \Om(u)=1+\frac 1{|a_1|\sqrt{u^2-a_1}}+\sum_{i=2}\frac{\xi_i}{|a_i|\sqrt{u^2-a_i}}.\]
It follows that $\Om(u)$ decreases from $\Om(0)=\mu(0)$ to $\Om(+\nf)=1$ and never vanishes, showing by the same argument displayed above that $M\cong {\mb H}^2$.

\nin The checks that $Q_1$ and $Q_2$ are globally defined are again elementary.$\quad\Box$

Let us add:

\begin{nth} The SI systems ${\cal I}_{-\pm}$, corresponding to $F(a)=(a-a_1)$, equipped with cubic integrals, do generalize to $\dst F(a)=\prod_{i=1}^n(a-a_i)$ with integrals of degree $\ 2n+1\ $ (with $n\geq 2)$) which remains globally defined on $M\cong{\mb H}^2$ under the following restrictions
\beq\label{constr-}
\barr{llll}
{\cal I}_{-+}: \quad & 0<a_1<+1,\quad a_i>1 \qq \& \qq   \xi_i>0 \qq \& \qq\dst \frac 1{|a_1|^{3/2}}+\sum_{i=2}^n\frac{\xi_i}{|a_i|^{3/2}}>1,\\[4mm]
{\cal I}_{--}: \quad & 0<a_1<+1,\quad a_i>1 \qq\& \qq \xi_i>0. \qq \earr\eeq
\end{nth}

\nin{\bf Proof:} The first system, above ${\cal I}_{-+}$,   is generated by
\[
x(a)=\frac a2-\frac 1{\sqrt{a_1-a}}-\sum_{i=2}^n\frac{\xi_i}{\sqrt{a_i-a}}\quad\Longrightarrow\quad  \mu(u)=1-\frac 1{(a_1-u^2)^{3/2}}-\sum_{i=2}^n\frac{\xi_i}{(a_i-u^2)^{3/2}}.\]
The function $\mu$ is decreasing from 
\[\mu(0)=1-\frac 1{a_1^{3/2}}-\sum_{i=2}^n\frac{\xi_i}{a_i^{3/2}}<0\qq\longrightarrow\qq \mu(\sqrt{a_1})=-\nf.\]
So we can define
\[
t=u\,\Om(u)\qq\qq \Om(u)=1-\frac 1{a_1\sqrt{u^2-a_1}}-\sum_{i=2}\frac{\xi_i}{a_i\sqrt{u^2-a_i}}.\]
It follows that $\Om(u)$ decreases from $\Om(0)=\mu(0)<0$ to $\Om(\sqrt{a_1})=-\nf$ and never vanishes, showing by the same argument given above that $M\cong {\mb H}^2$.

The second system, above ${\cal I}_{-+}$, is generated by
\[
x(a)=\frac a2+\frac 1{\sqrt{a_1-a}}+\sum_{i=2}^n\frac{\xi_i}{\sqrt{a_i-a}}\quad\Longrightarrow\quad  \mu(u)=1+\frac 1{(a_1-u^2)^{3/2}}+\sum_{i=2}^n\frac{\xi_i}{(a_i-u^2)^{3/2}}.\]
The function $\mu$ is increasing from
\[
\mu(0)=1+\frac 1{a_1^{3/2}}+\sum_{i=2}^n\frac{\xi_i}{a_i^{3/2}}<0\qq\longrightarrow\qq \mu(\sqrt{a_1})=+\nf.\]
So we can define
\[
t=u\,\Om(u)\qq\qq \Om(u)=1+\frac 1{a_1\sqrt{a_1-u^2}}+\sum_{i=2}\frac{\xi_i}{a_i\sqrt{a_i-u^2}}.\]
It follows that $\Om(u)$ increases from $\Om(0)=\mu(0)>0$ to $\Om(\sqrt{a_1})=+\nf$ and never vanishes, showing by the same argument as above that $M\cong {\mb H}^2$.

The checks that the integrals are globally defined are again elementary. $\quad\Box$

Let us give another example which is a close cousin of the system considered in Proposition \ref{R2}:

\begin{nth}\label{R4} For $n\geq 2$ the choice
\[
F(a)=(a-a_1)(a-a_2)\wh{F}(a) \qq\qq 0<a_1<a<a_2\]
with \footnote{If $n=2$ we have $\wh{F}(a)=1$.}
\[\wh{F}(a)=\prod_{i=3}^n(a-a_i): \qq \qq \Big(\ a_i<a_1\ \vee \  a_i>a_2 \qq i=3,\ldots n\ \Big)\]
leads to a SI system with the metric
\beq
g=\frac 1{A(t)}(dt^2+dy^2)\qq\qq (t,y)\in{\mb R}^2\eeq
globally defined on the manifold $M\cong{\mb R}^2$ as well as the integrals $S_1$ and $S_2$.
\end{nth}

\nin{\bf Proof:} Let us consider \footnote{For $n=2$ the last sum is absent.}
\[
x(a)=\frac{\nu}{2}\,a-\frac{\wti{\xi}_1}{\sqrt{a-a_1}}+\frac{\wti{\xi}_2}{\sqrt{a_2-a}}-\sum_{i=3}^n\frac{\eps_i\,\wti{\xi}_i}{\sqrt{\eps_i(a-a_i)}}\]
with
\[
\wti{\xi}_1=\sqrt{a_2-a_1}\,a_1\,\xi_1\qq\qq \wti{\xi}_2=\sqrt{a_2+a_1}\,a_2\,\xi_2 \qq \qq \wti{\xi}_i=\sqrt{a_2-a_1}\,a_i\,\xi_i\]
and
\[
\eps_i=+1\quad a_i<a_1\qq \& \qq \eps_i=-1 \quad a_i>a_2.\]
The coordinate change
\[
a=a_1+(a_2-a_1)s^2 \qq s\equiv \sin\tht: \quad a\in(a_1,a_2)\,\leftrightarrow\,\tht\in (0,\pi/2)\]
gives
\[
x(\tht)=\frac{\nu}{2}(a_2-a_1)\,s^2-\frac{\xi_1\,a_1}{s}+\frac{\xi_2\,a_2}{\sqrt{1-s^2}}-\sum_{i=3}^n\,\frac{\eps_i a_i\xi_i}{\sqrt{\eps_i(\rho_i+s^2)}} \qq\qq \rho_i=\frac{a_1-a_i}{a_2-a_1}.\]
So differentiating we get
\[
D_{\tht}\,x=\nu(a_2-a_1)\,sc+\xi_1\frac{a_1\,c}{s^2}+\xi_2\frac{a_2\,s}{(1-s^2)}+\sum_{i=3}^n\,\frac{a_i\xi_i\, s\, c}{(\rho_i+s^2)^{3/2}}\]
This relation show that $D_{\tht}\,x>0$.

Defining $\dst dt=\frac{dx}{\sqrt{a}}$ one gets
\[
t(s)=\sqrt{a_1+(a_2-a_1)s^2}\left(\nu-\frac{\xi_1}{s}+\frac{\xi_2}{\sqrt{1-s^2}}-\sum_{i=3}^n\,\frac{\eps_i\xi_i}{\sqrt{\rho_i+s^2}}\right).\]
From now on the proof follows exactly the same steps as in the proof of Proposition \ref{R2}.$\quad\Box$

\section{Conclusion}
Needless to say a lot of work is still necessary in the  study of the models constructed here. Let us just mention a few items:
\brm
\item Determine the integrals for the non simple cases. 
\item Find more globally defined systems with ${\mb H}^2$ or ${\mb R}^2$ for manifolds. It is possible that a general proof can be given that in this class of models (called ``affine case" in \cite{vds}) the possible manifolds are restricted to ${\mb H}^2$ or ${\mb R}^2$.
\item When the integrals are quadratic in the momenta we are back to a metric due to Koenigs. In this case it was shown in \cite{kkmw} that the integrals generate a quadratic algebra. Is there a generalization for integrals of higher degrees?
\item The integrable metrics shown to exist by Kiyohara in \cite{Ki} are Zoll metrics, which means that all the geodesics are closed. This suggests the conjecture that  the corresponding systems are in fact SI. Could it be proved?
\item An even more difficult task would be to check whether the classical integrability survives to quantization in the sense of \cite{dv}. 
\erm
To end up the author is somewhat disappointed because the manifolds obtained here never meet ${\mb S}^2$ and are of no use with respect to the Conjecture quoted in the Introduction.  However, it was shown in \cite{vds} that the ``hyperbolic case" is much more interesting since one is led either to teardrop orbifolds (Tannery orbifolds) or  ${\mb S}^2$. Unfortunately the analysis went through explicitly for cubic integrals but the generalization to higher degrees remains unknown, indicating a more difficult ground. However the reward could be here plenty of Zoll metrics (as is already the case for cubic integrals as shown in \cite{VaZ}) which are quite interesting geometric objects.

\begin{appendices}

\section{The cases of a non simple $F$}

When $F$ is simple, the general solution of the ODE (\ref{odeP3}) was given in Proposition \ref{propode}. This appendix will deal with all the remaining cases: either $F$ has a multiple real zero or it has multiple couples of complex conjugate zeroes. 

\subsection{A multiple real zero}
Here we have \footnote{If $r=n$ we take $\wh{F}=1$.}
\beq
F(a)=(a-a_1)^r\,\wh{F}(a)\qq\qq 2 \leq r \leq n \eeq 
where $\wh{F}$ is simple.

One has:
\begin{nth}\label{rmult}
For $\ F=(a-a_1)^r\,\wh{F}(a)$ with $\ 2\leq r\leq n$ and a generic $\wh{F}$, the solution 
of (\ref{odeP3}) is given by \footnote{If $r=n$ the sum over $i$ disappears.}
\beq
x=\sum_{k=1}^{r} \frac{\mu_k}{(\De_1)^{k-1/2}}+\sum_{i=r+1}^n\frac{\xi_i}{\sqrt{\De_i}}
\qq\qq \De_i=\eps_i(a-a_i).\eeq
The general solution of (\ref{solO}) is simply obtained by adding to $x(a)$ the linear term $\dst\frac{\nu_n}{2}\,a$. \end{nth}

\vspace{3mm}
\nin{\bf Proof:} Due to linearity we first check the ODE for
\[x_k(a)=[\eps_1(a-a_1)]^{-k+1/2}\qq\qq k\in\,
{\cal S}^1_k.\]
Since the variable is $a$ we will denote the derivation order by a superscript. Using (\ref{idop}) and interchanging the summations order we have first
\[{\rm Op}_n[(a-a_0)^r\,\wh{F}]x_k=\sum_{l=0}^n\frac{\wh{F}^{(n-l)}}{(n-l)!}{\rm Op}_l[(a-a_0)^r]x_k\]
and we will show that
\[\forall l \in{\cal S}_n^0\qq\forall k \in {\cal S}^1_r\qq X_{l,k}^r\equiv {\rm Op}_l[(a-a_0)^r]x_k=0.\]
Computing the various derivatives and defining $\si=s+r-l$ one gets
\beq\label{idhyper}
X_{l,k}^r=(-\eps_1)^{l-r}\,\De_1^{r-k-l+1/2}\,\frac{\G(k+l-r-1/2)}{\G(k-r+1/2)}\sum_{\si=0}^r\frac{(-1)^{\si}\,r!}{\si!\,(r-\si)!}\frac{(k-1/2)_{\si+l-r}}{(1/2)_{\si+l-r}}.
\eeq
Using the identities 
\beq\label{idPoch}
(a)_{N+\si}=(a)_N\,(a+N)_{\si}\qq (-r)_{\si}=(-1)^{\si}\,\frac{r!}{(r-\si)!}\quad \si\leq r 
\qq (-r)_{\si}=0 \quad r>\si \eeq
shows that the sum is proportional to 
\beq
\sum_{\si=0}^{\nf}\frac{(-r)_{\si}\,(k+l-r-1/2)_{\si}}{\si!\,(l-r+1/2)_{\si}}\eeq
which is a hypergeometric function \cite{emot}[p. 61] at unity 
\[_2F_1\left(\barr{c} -r,\ k+l-r-1/2\\l-r+1/2\earr\ ;\,1\right)=
\frac{\G(l-r+1/2)\,\G(r-k+1)}{\G(l+1/2)\,\G(1-k)}\]
and does vanish for all $k\in\,{\cal S}^1_r$. 

For the proof to be complete let us check now that
$\dst x_i=\frac 1{\sqrt{\eps_i(a-a_i)}}$ 
is also a solution of the ODE (\ref{odeP3}).

We start from
\[
{\rm Op}_n[(a-a_1)^r\wh{F}]x_i=\sum_{k=0}^n\frac{[(a-a_1)^r]^{(n-k)}}{(n-k)!}\,{\rm Op}_k[\wh{F}]x_i\]
and use relation (\ref{allk}) which states that
\[
\forall k\,\in\,{\cal S}^0_n: \qq {\rm Op}_k[\wh{F}]x_i=\frac{\sqrt{\De_i}}{k!}\,D_a^k\left(\frac{\wh{F}}{\De_i}\right).\]
Inserting this second relation into the first one and using Leibnitz fromula we conclude to
\[
{\rm Op}_n[(a-a_1)^r\wh{F}]x_i=\sqrt{\De_i}\,D_a^n\left(\frac{\wh{F}}{\De_i}\right)\]
which does vanish. $\quad\Box$

\vspace{3mm}
\nin{\bf Remarks:} 
\brm
\item Due to the linearity of the ODE for $x$ one can easily obtain its form in the generalized case where 
\[
F=(a-a_1)^{r_1}\cdots(a-a_s)^{r_s}\prod_{i=r+1}^n(a-a_i) \qq\qq r=r_1+\cdots r_s.\]
\item Let us observe that the solution obtained remains valid even if $a_1$ becomes complex.
\erm

\subsection{Multiple complex conjugate zeroes}
In this case we have:

\begin{nth}If $F$ has the structure
\[F(a)=(a^2+a_1^2)^r\,\wh{F}(a)\qq\qq 2\leq 2r \leq n \]
where $\wh{F}$ is simple, then the solution of the ODE (\ref{odeP3}) becomes \footnote{If $2r=n$ the second sum vanishes.}
\beq
x=\sum_{k=1}^{r}\Big(\mu_k^{+}\,P_k+\mu_k^{-}\,Q_k\Big)+\sum_{k=2r+1}^n\,\frac{\xi_i}{\sqrt{\De_i}}.
\eeq 
Defining
\beq
a=a_1\,\tan\tht\qq\qq \tht\in\,(-\frac{\pi}{2},+\frac{\pi}{2})\qq\qq a_0>0\eeq
we have
\beq\label{cmult}
\left\{\barr{l} \dst 
P_k=\cos\frac{\tht}{2}\ {\cal P}_k\\ [4mm] \dst Q_k=\sin\frac{\tht}{2}\,{\cal P}_k\earr\right. \qq \longrightarrow\qq {\cal P}_k=(\cos\tht)^{k-1/2}\Big[U_{k-1}(\cos\tht)-U_{k-2}(\cos\tht)\Big]
\eeq
where the $U_n$ are the Tchebyshev polynomials of second kind supplemented with $U_{-1}=0$.

\nin The general solution of (\ref{solO}) is again  obtained by adding to $x(a)$ the linear term $\dst\frac{\nu_n}{2}\,a$.\end{nth}

\vspace{3mm}
\nin{\bf Proofs:} Using the results of the previous proposition we know that if we start from
\[F(a)=(a-ia_1)^r(a+ia_1)^r\,\wh{F}(a)\qq\qq 2\leq 2r\leq n\]
the general solution is given by
\[x(a)=\sum_{k=1}^r\Big(\frac{\la_k^+}{(a-ia_1)^{k-1/2}}+\frac{\la_k^-}{(a+ia_1)^{k-1/2}}\Big)+\sum_{i=2r+1}^n\frac{\xi_i}{\sqrt{\De_i}}.\]
It is more convenient to write the first piece in $x$ as
\[\sum_{k=1}^r\Big(\frac{\mu_k^+}{(a_1+ia)^{k-1/2}}+\frac{\mu_k^-}{(a_1-ia)^{k-1/2}}\Big)\]
so that the basis required is just made out of the real and imaginary parts of $(a_1+ia)^{-k+1/2}$. This is most easily computed using
\[a=a_1\,\tan\tht\qq\qq \tht\in\,(-\frac{\pi}{2},\frac{\pi}{2})\qq\qq a_1>0.\]
We have
\[(1+ia/a_1)^{-k+1/2}=(\cos\tht)^{k-1/2}\,e^{-i(k-1/2)\tht}\]
which gives a first expression for the real and imaginary parts:
\[
P_k=(\cos\tht)^{k-1/2}\ \cos\Big((k-1/2)\tht\Big) \qq 
Q_k=(\cos\tht)^{k-1/2}\ \sin\Big((k-1/2)\tht\Big).\]
Recalling the defining relations of Tchebychev polynomials 
\[
T_n(\cos\tht)=\cos(n\tht)\qq\qq U_n(\cos\tht)=\frac{\sin(n+1)\tht}{\sin\tht}\]
we have
\[
P_k=(\cos\tht)^{k-1/2}\,\cos\frac{\tht}{2}\Big[T_k(\cos\tht)+(1-\cos\tht)U_{k-1}(\cos\tht)\Big]. \]
Using the relation \footnote{It is valid also for $k=1$ due to our convention that $U_{-1}=0$.}
\[
T_k(\cos\tht)-\cos\tht\,U_{k-1}(\cos\tht)=-U_{k-2}(\cos\tht) \]
gives for $P_k$ the formula (\ref{cmult}). A similar computation gives the required formula also for $Q_k$. 

Switching back to the variable $a$ we have
\[
P_k=\sqrt{\sqrt{a^2+a_1^2}+a_1}\,{\cal R}_k \qq\qq  
Q_k=\frac a{\sqrt{\sqrt{a^2+a_1^2}+a_1}}\,{\cal R}_k\]
with
\[
\forall k\in\,{\cal S}^1_r \qq\qq 
{\cal R}_k=(a^2+a_1^2)^{-k/2}
\Big[U_{k-1}\Big(\frac{a_1}{\sqrt{a^2+a_1^2}}\Big)-U_{k-2}\Big(\frac{a_1}{\sqrt{a^2+a_1^2}}\Big)\Big].
\]
For $k=1$ we have merely
\beq 
P_1(a)=\frac{\sqrt{\sqrt{a^2+a_1^2}+a_1}}{\sqrt{a^2+a_1^2}}\qq\qq Q_1(a)=\frac{a}{\sqrt{a^2+a_1^2}\,\sqrt{\sqrt{a^2+a_1^2}+a_1}}.
\eeq
 
Let us give a second proof, which does not use the complexification argument given above, and which is  similar to the proof given for Proposition \ref{rmult}. Setting $a_1=1$ we have to show that the function 
$\,x_k(a)=(1+ia)^{-k+1/2}$ is indeed a solution of the ODE (\ref{odeP3}) when $F=(a^2+1)^r\,\wh{F}$. 

Using twice the formula (\ref{idop}) we have
\[{\rm Op}_n\Big((a^2+1)^r\wh{F}\Big)x_k=\sum_{s=0}^n\frac{\wh{F}^{n-s}}{(n-s)!}
\sum_{l=0}^{s}\frac{[(1-ia)^r]^{(s-l)}}{(s-l)!}\,{\rm Op}_l((1+ia)^r)\,x_k\]
and we will prove that
\[\forall l\in{\cal S}_n^0 \qq \forall k\in\,{\cal S}^1_r\qq  X_{l,k}^r\equiv {\rm Op}_l((1+ia)^r)\,x_k=0.\]
Computing the various derivatives one gets
\[X_{l,k}^r\propto \sum_{s=l-r}^l\frac{(-1)^s\,r!}{(l-s)!\,(r-l+s)!}\frac{(k-1/2)_s}{(1/2)_s}
\propto \sum_{\si=0}^r\frac{(-1)^{\si}\,r!}{\si! (r-\si)!}\frac{(k-1/2)_{\si+l-r}}{(1/2)_{\si+l-r}}.\]
This sum was already found in (\ref{idhyper}) and shown to vanish for all $k\in\,{\cal S}^1_r$. $\quad\Box$

\section{Symmetric functions of the roots}
Let us take a set of numbers $\{a_1,a_2,\ldots,a_n\}$ and let us define
\beq\label{sfr0} 
P\equiv \prod_{k=1}^n\,(a-a_k)=\sum_{k=0}^n\,(-1)^k\,\si_k\,a^{n-k},
\eeq
where the $\si_k$ are the symmetric functions of the roots for the monic polynomial $P$.

Restricting ourselves to real polynomials $P$ this definition remains valid either if some root is multiple or if some complex conjugate couple of roots appear. 

It follows that
\beq\label{sik} 
F=\prod_{k=1}^n\,(a-a_k)=\sum_{k=0}^n A_k\,a^k\quad\Longrightarrow\quad  
(-1)^{k}\,\si_k=A_{n-k}\qq \forall k\in {\cal S}_n^0.
\eeq

In all what follows it will be supposed that $P$ is simple, which means that all the zeroes $a_i$ are simple.
So we can define
\beq\label{sfr1}   
\frac P{a-a_i}=\sum_{k=0}^{n+1}\,(-1)^{k-1}\,\si^i_{k-1}\,a^{n-k}\qq\Longrightarrow\qq\si^i_{-1}=\si^i_n=0.
\eeq
Multiplying this last relation by $\,a-a_i$ we get
\beq\label{id1sfr} 
\forall k \in {\cal S}_n^0: \qq\quad \si_k=\si^i_k+a_i\,\si^i_{k-1}.\eeq
An easy recurrence, using (\ref{id1sfr}) and (\ref{sik}), gives the relations
\beq\label{siik}
\forall k \in {\cal S}_n^1: \qq \si^i_{k-1}=\sum_{s=0}^{k-1}(-a_i)^{k-1-s}\,\si_s=(-1)^{k-1}\sum_{s=0}^{k-1} a_i^{k-1-s}\,A_{n-s}.
\eeq

For $i\neq j$ we can define
\beq\label{sfr2} 
\frac P{(a-a_i)(a-a_j)}=\sum_{k=0}^{n+2}\,(-1)^{k-2}\,\si^{ij}_{k-2}\,a^{n-k} \ \Longrightarrow\  \si^{ij}_{-2}=\si^{ij}_{-1}=\si^{ij}_{n-1}=\si^{ij}_n=0.
\eeq
Multiplying this last relation by $\,(a-a_j)\,$ we get 
\beq\label{id2sfr}
\forall k\in {\cal S}^0_n: \qq\qq \si^i_{k-1}=\si^{ij}_{k-1}+a_j\,\si^{ij}_{k-2} 
\eeq
from which we deduce
\beq\label{id3sfr}
\forall k\in {\cal S}^0_n \qq\qq  \si^{ij}_{k-2}=-\,\frac{\si^i_{k-1}-\si^j_{k-1}}{a_i-a_j}.
\eeq
A quadratic identity follows from the relation
\beq
i\neq j: \quad \frac P{a-a_i}\frac P{a-a_j}=P\frac P{(a-a_i)(a-a_j)}.\eeq
As a preliminary remark let us observe that any product of the form
\[
AB=\sum_{k,l=1}^n\,A_k\,B_l\,H^{2n-k-l}\,P_y^{2(k+l)}\]
after setting $s=k+l$ and reversing the order of the summations becomes
\[
\sum_{s=2}^n\,U_s(A,B)\,H^{2n-s}\,P_y^{2s}+\sum_{s=n+1}^{2n}\,V_s(A,B)\,H^{2n-s}\,P_y^{2s}.\]
with
\[
U_s(A,B)=\sum_{k=1}^{s-1}\,A_k\,B_{s-k}\qq\qq 
V_s=\sum_{k=s-n}^n\,A_k\,B_{s-k}.\]
Taking this observation into account one obtains the relations
\beq\label{idQ}
\barr{lcll}\dst 
\sum_{k=1}^{s-1}\,\si^i_{k-1}\,\si^j_{s-k-1} & = & \dst \sum_{k=1}^{s-1}\,\si_k\,\si^{ij}_{s-k-2}+\si^{ij}_{s-2} & \qq s\in\{2,3,\ldots,n\}\\[6mm]\dst 
\sum_{k=s-n}^{n}\,\si^i_{k-1}\,\si^j_{s-k-1} & = & \dst \sum_{k=s-n}^{n}\,\si_k\,\si^{ij}_{s-k-2} & \qq s\in\{n+1,\ldots,2n\}.\earr
\eeq

To conclude, the symmetric functions needed  by our analysis are therefore
\beq\barr{ccccccc}
\Big(\si_0=1 & \si_1 & \si_2 & \ \ldots\ &\si_{n-2} & \si_{n-1} & \si_n\Big)\\[6mm]
\Big(\si^i_0=1 & \si^i_1 & \si^i_2 & \ \ldots\ &\si^i_{n-2} & \si^i_{n-1}\Big) & \si^i_n=0\\[6mm]
\Big(\si^{ij}_0=1 & \si^{ij}_1 & \si^{ij}_2 & \ \ldots\ &\si^{ij}_{n-2}\Big) & \si^{ij}_{n-1}=0 & \si^{ij}_n=0
\earr\eeq
\end{appendices}

\end{document}